%% file: main.tex
\documentclass{article} 
\usepackage{iclr2027_conference,times}
\renewcommand{\headrulewidth}{0pt}
\input{math_commands.tex}

\usepackage{hyperref}
\usepackage{url}
\usepackage{booktabs}
\usepackage{graphicx}
\usepackage{subcaption}
\usepackage{titletoc}
\usepackage{longtable}

\title{Learning Spectrally Optimised Mesh-Free \\ Discretisations}

\author{
Lucas Gerken Starepravo$^{1}$, Henry Broadley$^{2}$, Steven Lind$^{3}$ \& Jack R. C. King$^{1}$ \\
$^{1}$School of Engineering, University of Manchester, Manchester, United Kingdom \\
$^{2}$School of Engineering, University of Liverpool, Liverpool, United Kingdom \\
$^{3}$School of Engineering, Cardiff University, Cardiff, United Kingdom \\
\texttt{lucas.gerkenstarepravo@postgrad.manchester.ac.uk}
}

\iclrfinalcopy 
\begin{document}

\maketitle

\begin{abstract}
Numerical methods for partial differential equations (PDEs) discretise differential operators, ideally reproducing the action of the continuous operators across all wavenumbers permitted by a given discretisation. Spectral-type methods approach this ideal, but rely on structured grids or high-order meshes that are hard to generate for complex geometries. In contrast, mesh-free methods are geometrically flexible, yet how faithfully they reproduce the operator across resolved scales is strongly influenced by a heuristically chosen kernel, which selects one of many weight sets satisfying the same consistency conditions without regard to the resulting spectral response. To address this, we introduce Spectrally optimised Neural Discretisations (SpeND), a framework that learns the map from local stencil geometry to discretisation weights on unstructured point clouds. A projection layer constrains every predicted stencil to the affine set defined by the discrete moment conditions, so polynomial consistency, and hence formal order of accuracy, holds exactly. Since consistency is enforced by the architecture, the weights can additionally be optimised for accuracy on a prescribed function space. Herein, we use Fourier modes and target the exact differentiation response over a chosen wavenumber band, so training is unsupervised. The loss function is a design interface: changing how it weights wavenumbers yields operators with distinct accuracy profiles. The trained discrete differential operators are PDE-agnostic and are applied without retraining to the Poisson, Burgers and Navier--Stokes equations, to near-boundary stencils, and across resolutions. At the same order and resolution, SpeND matches or improves on established mesh-free discretisations, and has been shown to reduce the wall-clock time between $3-20\times$ to achieve equivalent error as the baselines.
\end{abstract}

\section{Introduction}

Partial Differential Equations (PDEs) play an essential role in science and engineering \citep{stocker2013climate, MARTINEZ2017654, Johnston2006NonNewtonian}, however, analytical solutions are rarely available for problems of practical interest. Consequently, numerical methods are the standard tools for high-fidelity simulation, data generation, and the study of new physical regimes. Their reliability rests on a guarantee structure that is independent of the PDE being solved: a discretisation that is consistent, meaning it approximates the continuous operator, and stable, meaning errors remain bounded, is convergent. For well-posed linear problems this relationship is formalised by the Lax equivalence theorem~\citep{lax_survey}.

Global spectral methods~\citep{Orszag1971, GottliebOrszag1977} sit at the accuracy ceiling of numerical methods: the symbol of the discrete operator (i.e. its action on each Fourier mode) matches that of the continuous one across the wavelengths permitted by the discretisation, so a target accuracy is reached with far fewer degrees of freedom than any fixed-order scheme. This accuracy is conditional on structured collocation points, which restricts global spectral methods to simple geometries~\citep{Boyd2001}. Domain decomposition approaches such as spectral element methods relax this constraint by partitioning the domain into elements, at the cost of the global symbol match. Geometric flexibility is then limited by mesh generation, where high-order meshes remain difficult to produce for complex geometries~\citep{KarniadakisSherwin2005}.

Unstructured strong-form methods such as Smoothed Particle Hydrodynamics (SPH)~\citep{Lucy1977}, radial basis function-finite difference methods (RBF-FD)~\citep{tolstykh2000using,Bayona2013Flames,Bayona2015}, and the local anisotropic basis function method (LABFM)~\citep{King2020} approximate discrete differential operators solely based on local neighbour information and through a compactly supported kernel. The kernel, however, is chosen heuristically, so how faithfully the discrete operator reproduces the continuous one across resolvable scales is fixed by a modelling choice made that offers little control of that objective.

Machine learning (ML) has been applied at several levels of the PDE solution pipeline. Neural operators and learned surrogates replace the solver with a map between function spaces~\citep{Lu2021, Li2021FNO, Anandkumar2019GKN}, achieving large speedups but yielding a model tied to the family of PDEs it was trained on, and without the convergence guarantees of classical numerical methods. Closer to the present work, learned discretisations retain the numerical solver and learn only the stencil coefficients~\citep{BarSinai2019, Kochkov2021}, recovering coarse-grid accuracy while preserving the solver structure. These methods, however, are restricted to structured grids and are trained with supervision against high-resolution reference solutions, which ties the learnt discretisation to flows with similar local structures.


In this context, we introduce Spectrally optimised Neural Discretisations (SpeND), a framework in which a neural network predicts mesh-free stencil weights on unstructured point clouds. The framework has three key properties. (i) \textbf{Guaranteed consistency order.} A projection layer constrains the network output to the affine set satisfying the discrete moment conditions, so the target consistency order holds exactly for every predicted stencil. (ii) \textbf{Numerically grounded training.} Training minimises the deviation between the analytically known modal response of the continuous and the discretised operator over a prescribed band, yielding an operator that is accurate across the desired spectrum rather than for a particular flow or PDE. (iii) \textbf{Generalisation.} A single trained operator transfers without retuning across resolutions and PDEs.



\section{Background \& Related Work}
\subsection{Numerical Approximation of Differential Operators}
\label{sec:num_approx}

Let the spatial domain $\Omega \subset \mathbb{R}^d$ be represented by a set of collocation points $\mathcal{P} := \{\mathbf{x}_i\}_{i=1}^P \subset \Omega$, where $\mathbf{x}_i$ denotes the position of node $i$. For a continuous differential operator $D$ acting on a field $\phi$, a local discrete approximation at node $i$ takes the form\footnote{A full list of symbols is given in the \hyperref[app:nomenclature]{Nomenclature}.}
\begin{equation}
\label{eq:discrete_op}
    L^D[\phi(\mathbf{x}_i)] = \sum_{j \in \mathcal{N}_i} \phi_{ji} \, w_{j,i}^D,
\end{equation}
in which $L^D$ is the discrete counterpart of $D$, the index set $\mathcal{N}_i$ collects the neighbours forming the computational stencil of node $i$, each $w_{j,i}^D$ is the weight assigned to neighbour $j$ within that stencil, and differences are written $(\cdot)_{ji} := (\cdot)_j - (\cdot)_i$. With suitable choices of stencil and indexing, \eqref{eq:discrete_op} recovers a wide range of numerical methods, among them finite difference (FD)~\citep{5391985}, finite elements (FE)~\citep{Clough1990} and SPH. What distinguishes one discretisation from another is the construction of the neighbourhoods $\mathcal{N}_i$ and of the weights $w_{j,i}^D$.

Consistency characterises how the discrete operator approaches the continuous one as the node spacing $s$ tends to zero. A discrete operator $L^D$ is consistent to order $p$ if it reproduces every Taylor coefficient of $D$ up to polynomial degree $p$, so that, for an operator $D$ of order $m$, the truncation error is $\mathcal{O}(s^{p+1-m})$. This can be written as a linear system for the stencil weights. Collecting the stencil weights of node $i$ into the vector $\mathbf{w}_i^D := (w_{j,i}^D)_{j \in \mathcal{N}_i}$, substituting the multivariate Taylor expansion of $\phi_{ji}$ about $\mathbf{x}_i$ into~\eqref{eq:discrete_op} and matching the discrete and continuous operators term by term yields the linear moment system
\begin{equation}
\label{eq:moment_system}
    \mathbf{V}_i \, \mathbf{w}_i^D = \mathbf{d}^D, \qquad [\mathbf{V}_i]_{\boldsymbol{\alpha}, j} = \frac{\mathbf{x}_{ji}^{\boldsymbol{\alpha}}}{\boldsymbol{\alpha}!}, \qquad [\mathbf{d}^D]_{\boldsymbol{\alpha}} = \frac{D\,\mathbf{x}^{\boldsymbol{\alpha}}\big|_{\mathbf{x}=\mathbf{0}}}{\boldsymbol{\alpha}!}.
\end{equation}
Rows are indexed by the $N_p = \binom{p+d}{d} - 1$ multi-indices $\boldsymbol{\alpha}$ with $1 \leq |\boldsymbol{\alpha}| \leq p$, one per Taylor monomial, and columns by the neighbours $j \in \mathcal{N}_i$, so that $\mathbf{V}_i \in \mathbb{R}^{N_p \times |\mathcal{N}_i|}$ and $\mathbf{w}_i^D \in \mathbb{R}^{|\mathcal{N}_i|}$. The entries of the stencil moment vector $\mathbf{d}^D$ record the action of $D$ on each monomial, so the stencil is required to reproduce those coefficients exactly and to eliminate all remaining ones up to degree $p$. For instance, taking $d = 2$, $p = 2$ and ordering $\boldsymbol{\alpha}$ as $(1,0), (0,1), (2,0), (1,1), (0,2)$ gives $\mathbf{d}^{\partial_x} = (1,0,0,0,0)^\top$ for the first derivative and $\mathbf{d}^{\nabla^2} = (0,0,1,0,1)^\top$ for the Laplacian.

When $N_p = |\mathcal{N}_i|$ and $\mathbf{V}_i$ is non-singular, the moment system admits a unique solution. On a structured grid this recovers the finite difference method. Mesh-free discretisations on unstructured point clouds instead operate with $N_p < |\mathcal{N}_i|$ to improve the conditioning of $\mathbf{V}_i$ under irregular node spacing, so the system is underdetermined and consistency alone does not determine the weights.

\subsection{Classical Solvers}
\paragraph{Spectral, spectral-like, and spectrally-optimised frameworks.} Global spectral methods make differentiation exact for the resolvable band allowed by the grid and converge exponentially, but they rely on a structured tensor-product grid and on boundary conditions compatible with the basis \citep{Boyd2001}. Spectral element methods recover geometric flexibility by partitioning the domain into elements carrying local high-order bases \citep{KarniadakisSherwin2005}, yet the resolved band is no longer represented exactly, and the required high-order mesh is often the bottleneck for such schemes. Global radial basis function collocation removes the mesh entirely and retains near-spectral accuracy on scattered nodes, at the cost of dense operators, ill-conditioning that worsens as accuracy improves, and $\mathcal{O}(P^2)$ operations per time step \citep{Fornberg2011}.

Compact schemes abandon the exact symbol and instead match the modified wavenumber of a narrow stencil to the exact one over a prescribed band, achieving spectral-like resolution at low bandwidth \citep{Lele1992}, an idea since extended to scattered nodes by compact moving least squares~\citep{TRASK2016596} and compact LABFM~\citep{BROADLEY2026115308}. In all cases the derivative is defined implicitly, requiring a sparse global solve at every evaluation, which couples the whole domain and removes the locality that makes explicit stencils cheap to refine and parallelise.

Closest to our objective are explicit schemes whose coefficients are chosen to minimise modal error rather than to maximise formal order \citep{TAM1993262}, though the construction is restricted to structured stencils. The multi-kernel approach~\citep{broadley2025} also targets the modal response, but operates at a different level; it combines several instances of a given mesh-free discretisation. SpeND acts on the discretisation itself and could equally serve as the base scheme in such a combination.

\textbf{Local mesh-free discretisation methods} construct each differential operator by summation over a compact stencil of scattered neighbours, requiring no connectivity. This makes them geometrically flexible, tolerant of large deformation, and highly parallelisable, since every stencil is built independently. SPH~\citep{Lucy1977} was among the first such methods, but its weights derive from a fixed radial smoothing kernel and it is at best zeroth order consistent. A broader class of local methods instead imposes the moment conditions of \eqref{eq:moment_system} directly when constructing the weights. Notable examples are the generalised finite differences method (GFDM)~\citep{JENSEN197217}, generalised moving least squares (GMLS)~\citep{gmls_trask}, RBF-FD~\citep{tolstykh2000using} and LABFM~\citep{King2020}.

In all the mesh-free methods mentioned above, the weights are built from a heuristic radial function whose properties can significantly affect the solution quality (and consequently how well the discrete operator approximates the resolvable band)~\citep{dehnen_aly,LeBorne2023}.  Because the response cannot be prescribed, accuracy over the resolvable band can only be improved indirectly. Raising the order of consistency improves agreement between the discrete and continuous operators in the low-wavenumber limit, at the cost of $N_p$ growing combinatorially with $p$ and $d$, while refining the node spacing lowers the normalised wavenumbers at which the solution must be resolved, at a cost of $\mathcal{O}(s^{-d})$ nodes. Neither acts on the response across the band, leaving the stencil freedom admitted by~\eqref{eq:moment_system} unexploited.


\subsection{Neural solvers}

Many approaches have been proposed for solving PDEs with machine learning. Physics-informed neural networks~\citep{Raissi2019PINN} fit the solution of a single PDE instance by minimising its residual, and in their standard form must be retrained for each new instance; they also inherit the spectral bias of the network, which limits them on multiscale problems~\citep{WANG2022110768,rahaman2019spectralbiasneuralnetworks}. Neural operators~\citep{Lu2021,Li2021FNO} instead learn a mapping between function spaces from example solutions, and are correspondingly tied to the equation family and boundary conditions on which they were trained. Both approaches have been applied successfully across a wide range of PDEs, but physical consistency and stability are properties of a trained model on a particular problem, assessed empirically rather than guaranteed by construction. 

\paragraph{Hybrid solvers.} A second line preserves the classical solver and learns components within it~\citep{greenfeld2019,satorras2019,hsieh2019,um2021,brandstetter2023}, targeting problems from the acceleration of iterative solvers to the stability of autoregressive rollouts. Closest to this work are \textbf{learned discretisations}. \citet{BarSinai2019} learn finite volume stencil weights on coarse Cartesian grids, predicting them from the local field values with a convolutional network and training against the coarse-grained time derivative of a high-resolution reference. Consistency is imposed by construction, the prediction being a fixed low-order stencil plus a component in the null space of a truncated moment system. \citet{Zhuang2021,Kochkov2021,DEROMEMONT2026106978} extend this to more complex flows, and \citet{dresdner2023learningcorrectspectralmethods} learn corrections to a spectral solver on periodic grids. While effective, these methods are restricted to structured grids and are supervised on a high-resolution reference solution, which ties the learned discretisation to the equation and similar flow structures used to train it. The weights are also a function of the instantaneous field, so the network must be evaluated at every step of the simulation.


\paragraph{Beyond structured grids.} \citet{Choi2025Foundation} replace the classical polynomial bases of the finite element method with locally learned ones that transfer across geometries, mesh types and boundary conditions. The method remains mesh-based, requires data generated over many meshes and PDEs, and its convergence and stability remain an open problem. In the mesh-free setting, \citet{starepravo2026learningmeshfreediscretedifferential} learn discretisations by imposing the Taylor moment conditions as a loss penalty, removing the need for PDE data at training time. Thus, the resulting weights are only approximately consistent, with moment residuals well above machine precision, which restricts the framework to low orders of approximation. Furthermore, as in classical mesh-free schemes, the modal response is left to whatever the construction yields rather than being targeted.

\section{Method}
\subsection{Problem Setup}
\label{sec:setup}
Let $\mathcal{P}$ be an unstructured point cloud in $\Omega$. We associate to each point $\mathbf{x}_i$ the local neighbourhood $\mathcal{N}_i := \{j : \mathbf{x}_j \in \mathcal{P},\ \|\mathbf{x}_{ji}\|_2 \leq R_i\}$, where $R_i$ is the Euclidean distance from $\mathbf{x}_i$ to the farthest of its $N$ nearest nodes, $\mathbf{x}_i$ included, so that $|\mathcal{N}_i| = N$. Fixing the neighbour count rather than the support radius is common practice in mesh-free frameworks~\citep{BENITO20011039,Flyer2020}. Our objective is to construct, at each node, a discrete approximation $L^D$ of a differential operator $D$ acting on $\phi$, taking the form of the weighted sum~\eqref{eq:discrete_op} over $\mathcal{N}_i$. The approximation is therefore fully specified by the stencil weights $\mathbf{w}_i^D$.

In SpeND, we formulate the construction of the stencil weights as a learning problem. Rather than fixing the map from local geometry to weights through a kernel and a closure, we parametrise it by a neural network and constrain its output so that consistency holds exactly.

The network acts on the local geometry alone. Neighbour positions are normalised onto the unit circle $\bar{\mathbf{x}}_{ji}=\mathbf{x}_{ji}/R_i$, so the learned map depends only on the local node arrangement, and can be applied to different resolutions. A network $f_\theta$ maps the normalised positions to candidate weights $\tilde{\mathbf{w}}_i^D = f_\theta(\{\bar{\mathbf{x}}_{ji}\}_{j \in \mathcal{N}_i})$, which are then projected by $\Pi_i$ onto the affine set of solutions of the moment system $\bar{\mathbf{V}}_i \bar{\mathbf{w}}_i^D = \mathbf{d}^D$, i.e.\ \eqref{eq:moment_system} assembled from the normalised positions. The projection is affine and determined entirely by $\bar{\mathbf{V}}_i$, hence every stencil predicted by the network is consistent to order $p$ by construction. The projected weights are returned to the physical scale of the point cloud by a factor $R_i^{-m}$, giving
\begin{equation}
    \mathbf{w}_i^D = R_i^{-m}\,\Pi_i\big[f_\theta(\{\bar{\mathbf{x}}_{ji}\}_{j \in \mathcal{N}_i})\big].
\end{equation}
While the projection layer ensures the predicted weights are consistent, the system remains underdetermined for $N > N_p$. SpeND acts on this freedom, training $\theta$ so that the operator assembled from the projected weights optimises the modal response of $D$ across the resolved band.

\subsection{Consistency Constraints -- Moment Projection}
\label{sec:projection}
Since $N > N_p$ in the regime of interest, $\bar{\mathbf{V}}_i$ is wide and, provided it has full row rank, the consistent weights form an affine subspace of dimension $N - N_p$. We map the candidate weights $\tilde{\mathbf{w}}_i^D$ onto this subspace through the minimum-norm correction
\begin{equation}
\label{eq:projection}
    \bar{\mathbf{w}}^{D}_{i} = \Pi_i\big[\tilde{\mathbf{w}}_i^D\big] = \tilde{\mathbf{w}}_i^D - \bar{\mathbf{V}}_i^{+}\!\left(\bar{\mathbf{V}}_i\, \tilde{\mathbf{w}}_i^D - \mathbf{d}^D\right),
\end{equation}
where $\bar{\mathbf{V}}_i^{+}$ is the Moore--Penrose pseudoinverse. The correction vanishes whenever $\tilde{\mathbf{w}}_i^D$ is already consistent, so consistency of order $p$ holds for any network output and is a property of the architecture rather than of the converged parameters. Since the map is affine, its Jacobian is the orthogonal projector $\mathbf{P}_i = \mathbf{I} - \bar{\mathbf{V}}_i^{+}\bar{\mathbf{V}}_i$ onto $\ker\bar{\mathbf{V}}_i$, and differentiating through the layer requires no implicit solve. 

\subsection{Training}
\label{sec:training}
With consistency enforced by the projection, the objective determines which point of the consistent subspace is selected. Since both the network and the objective depend only on the normalised geometry, we write the objective in the normalised coordinates of a single stencil, with $\mathbf{x}_i$ at the origin. Let $\Phi$ be a class of test functions and $D$ the continuous operator that $L^D$ approximates. For each node $i$ and each $\phi \in \Phi$, the stencil residual is
\begin{equation}
  r_i[\phi] \;=\; \sum_{j \in \mathcal{N}_i} \phi_{ji}\, \bar{w}_{j,i}^D(\theta) \;-\; D[\phi](\mathbf{0}),
  \label{eq:stencil_residual}
\end{equation}
where $\bar{w}_{j,i}^D(\theta)$ are the projected network outputs, $\phi$ is evaluated at the normalised positions $\bar{\mathbf{x}}_{ji}$, and $D[\phi](\mathbf{0})$ is available analytically. No reference solutions are therefore required at any point in training. 

In this work we take $\Phi$ to be the Fourier modes $\phi_{\mathbf{k}}(\mathbf{x}) = e^{\mathrm{i}\mathbf{k}\cdot\mathbf{x}}$, with $\mathbf{k}$ measured in normalised coordinates, so that \eqref{eq:stencil_residual} measures the departure of the discrete operator from the exact continuous one at wavenumber $\mathbf{k}$. Optimising at a single wavenumber would leave the response at all others unconstrained. We therefore optimise the modal response over a band, $0 \leq \|\mathbf{k}\| \leq \eta\,k_{\text{Ny}}$, where $\eta \in (0,1]$ and $k_{\text{Ny}} = \pi/\bar{s}_N = \sqrt{\pi N}$ is the Nyquist wavenumber (i.e. the mininum wavenumber a given discretisation can resolve) at the mean normalised spacing $\bar{s}_N = \sqrt{\pi/N}$, obtained for $d = 2$ by distributing the $N$ stencil nodes uniformly over the unit disk.

The residual is a single complex number per mode. We make this concrete for $D = \partial_x$; the remaining operators follow analogously (Appendix~\ref{app:resolving}). The continuous operator returns $\partial_x e^{\mathrm{i}\mathbf{k}\cdot\mathbf{x}}\big|_{\mathbf{x}=\mathbf{0}}
= \mathrm{i}k_x$, while the discrete operator returns its modal response $\hat{L}^{\partial_x}_i(\mathbf{k}) := \sum_{j \in \mathcal{N}_i} \bar{w}^{\partial_x}_{j,i}\big(e^{\mathrm{i}\mathbf{k}\cdot\bar{\mathbf{x}}_{ji}} - 1\big)$. We then divide the discrete response by $\mathrm{i}$ and define the effective wavenumber $k_\text{eff}(\mathbf{k}) := -\mathrm{i}\,\hat{L}_i^{\partial_x}(\mathbf{k})$ (the stencil index is omitted for brevity), the wavenumber the discrete operator reproduces when differentiating $\phi_{\mathbf{k}}$. An exact discrete operator would give $k_\text{eff} = k_x$, so the deviation of $k_\text{eff}$ from $k_x$ measures the error of the discretisation at that mode. Separating real and imaginary parts,
\begin{equation}
\label{eq:keff}
    k_\text{eff}(\mathbf{k}) = \sum_{j \in \mathcal{N}_i} \bar{w}^{\partial_x}_{j,i}
    \sin(\mathbf{k} \cdot \bar{\mathbf{x}}_{ji})
    \;+\; \mathrm{i} \sum_{j \in \mathcal{N}_i} \bar{w}^{\partial_x}_{j,i}
    \left[ 1-\cos(\mathbf{k} \cdot \bar{\mathbf{x}}_{ji}) \right].
\end{equation}
The real part governs the phase speed of the mode, $\operatorname{Re}\{k_\text{eff}\}\operatorname{sign}(k_x) < |k_x|$ makes the mode lag the exact
solution and $\operatorname{Re}\{k_\text{eff}\}\operatorname{sign}(k_x) > |k_x|$ makes it lead, exceeding the physical propagation speed~\citep{soton23038}. The imaginary part governs the amplitude: for advection with positive velocity, $\operatorname{Im}\{k_\text{eff}\} < 0$ damps the mode, whereas $\operatorname{Im}\{k_\text{eff}\} > 0$ injects energy and usually has to be
suppressed by hyperviscous stabilisation~\citep{Jameson1981}; see Appendix~\ref{app:role}. 

Both parts of $k_\text{eff}$ vary continuously with wavenumber magnitude and propagation
angle, so we evaluate them on a finite set of modes. Let
$\mathcal{K} \subset \{\mathbf{k} : \|\mathbf{k}\| \le \eta k_{\text{Ny}},\, k_y \geq 0\}$ be a
fixed set of wavenumbers discretising the resolved half-disk. The omitted half is determined
by the parity of~\eqref{eq:keff},
$\operatorname{Re}\{k_\text{eff}(-\mathbf{k})\} = -\operatorname{Re}\{k_\text{eff}(\mathbf{k})\}$
and
$\operatorname{Im}\{k_\text{eff}(-\mathbf{k})\} = \operatorname{Im}\{k_\text{eff}(\mathbf{k})\}$,
so no information is lost. The per-stencil dispersion loss is
\begin{equation}
\label{eq:loss_disp}
\begin{split}
\mathcal{L}^{\text{disp}}_i &= \frac{1}{|\mathcal{K}|} \sum_{\mathbf{k} \in \mathcal{K}}
\sigma(\mathbf{k})\, \omega_>(\mathbf{k})
\left(\operatorname{Re}\{k_\text{eff}(\mathbf{k})\} - k_x\right)^2, \\
\omega_>(\mathbf{k}) &=
\begin{cases}
  \lambda_> & \operatorname{Re}\{k_\text{eff}(\mathbf{k})\}\,\operatorname{sign}(k_x) > |k_x|, \\
  1 & \text{otherwise},
\end{cases}
\end{split}
\end{equation}
with $\lambda_> > 1$ and $\sigma(\mathbf{k}) = \max(|k_x|/k_{\text{Ny}}, \epsilon)^{-2}$, normalised to unit
mean over $\mathcal{K}$. The two factors act independently; $\omega_>$ breaks the symmetry of the squared error, penalising $\lambda_>$ times more for a leading response, while $\sigma$ converts the objective from absolute to relative error, so that modes with a large exact response do not dominate the accuracy a solver depends on at small $k_x$. The floor $\epsilon$ bounds the weight where the exact response vanishes, avoiding blow-up as $|k_x|/k_{\text{Ny}} \to 0 $.

We penalise dissipation with the same asymmetric form, now penalising $\lambda_+$ times more for
amplification,
\begin{equation}
\label{eq:loss_diss}
\mathcal{L}^{\text{diss}}_i = \frac{1}{|\mathcal{K}|} \sum_{\mathbf{k} \in \mathcal{K}}
\omega_+(\mathbf{k})
\left(\operatorname{Im}\{k_\text{eff}(\mathbf{k})\}\right)^2,
\qquad
\omega_+(\mathbf{k}) =
\begin{cases}
  \lambda_+ & \operatorname{Im}\{k_\text{eff}(\mathbf{k})\} > 0, \\
  1 & \text{otherwise}.
\end{cases}
\end{equation}
The unweighted branch still penalises damping, since excess dissipation removes energy from resolved modes, but the asymmetry reflects that the two failures are not comparable: damping degrades accuracy, whereas amplification is a stability failure. We apply no relative weighting here, since the exact imaginary response is zero.

The network predicts stencil weights for arbitrary local geometries, so the objective is averaged over a corpus of stencils sampled from point clouds generated in advance
(Appendix~\ref{app:data}),
\begin{equation}
\label{eq:loss_total}
    \mathcal{L}(\theta) = \mathbb{E}_{i \sim \mathcal{S}} \left[ \mathcal{L}^{\text{disp}}_i
    + \gamma\, \mathcal{L}^{\text{diss}}_i \right],
\end{equation}
where $\mathcal{S}$ is the corpus of stencils and $\gamma > 0$ balances the two error types. Every term is computed from the predicted weights and the exact response of $D$, so training requires no PDE solutions at any point.

\subsection{Architecture}
The contribution of this work lies in the discretisation framework rather than the network, so the architecture is kept deliberately simple. We instantiate $f_\theta$ as an adapted PointNet \citep{pointnet} acting on each stencil independently, with shared per-point maps and a symmetric pool over the neighbours. The construction is permutation-equivariant, matching the absence of any physical meaning in the ordering of a neighbour list. Details are given in Appendix~\ref{app:architecture}.

\section{Experiments}
\label{sec:experiments}
We compare SpeND with RBF-FD \citep{tolstykh2000using,flyer2016polyI}, using polyharmonic splines with polynomial augmentation, and LABFM \citep{King2020}, using the Wendland C2 kernel with Hermite polynomials. Both are mature schemes that have been applied to complex flow problems \citep{rbf_fd1,bayona2017polyII,King2022,KING2024116762}, and both take the same inputs as SpeND, an unstructured mesh-free stencil, so no mesh or structured grid is assumed anywhere in the comparison. Learned solvers are frequently compared against baselines run at a different accuracy~\citep{McGreivy2024}. We therefore sweep the resolution of each method and report wall-clock time against error across the range, so the methods are compared at equal accuracy rather than at a single configuration, and hold the order of approximation and stencil size matched throughout, so that differences reflect only how each method spends the degrees of freedom left after consistency is imposed.

The experiments test whether a single trained model transfers, without retraining, across the settings a discretisation must handle: collocation point distributions, orders of approximation, and PDEs. All test cases are two-dimensional and admit analytical solutions, so measured error is attributable to the spatial discretisation rather than to a numerical reference carrying error of its own. Every stencil is limited to 30 neighbours, unless otherwise stated; the effect of stencil size is examined in Appendix~\ref{app:ablation}.


\subsection{Toy Problem}
\label{sec:toy}
We begin with a pure approximation test, which isolates the accuracy of the discrete operators from any solver. We evaluate six operators, pairing $\partial_x$ and $\nabla^2$ with consistency order $p \in \{2,3,4\}$. The $p = 2$ and $p = 4$ networks are trained. Since consistency is imposed by the projection rather than learned, the approximation order of a trained network can also be changed at inference, and to demonstrate this the $p = 3$ operators reuse the $p = 2$ network with only the moment constraints changed. The target function on the unit square $(x,y) \in [0,1]^2$ is a unit square wave in $x$, truncated to its first four non-zero harmonics,
\begin{equation}
\phi(x, y) = \frac{4}{\pi}\,\sin(2\pi y) \sum_{n=1}^{4}
\frac{\sin\!\big(2\pi(2n-1)(x - \tfrac{1}{4})\big)}{2n-1}.
    \label{eq:test_function}
\end{equation}
The function contains several discrete wavenumbers rather than a single one. The nodes are distributed in the domain with the particle shifting iterative procedure~\citep{FLYER201639}, which is standard in mesh-free simulations.

\begin{figure}[htbp]
\begin{center}
  \centering
\includegraphics[width=0.5\linewidth]{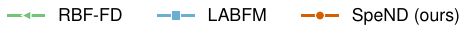}\\

\begin{subfigure}{0.16\linewidth}
  \centering
  \includegraphics[width=\linewidth]{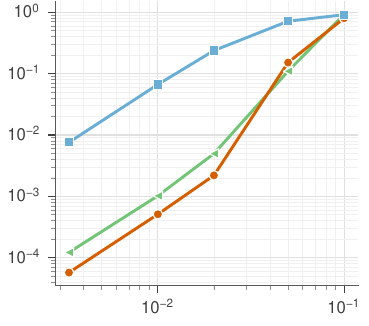}
\end{subfigure}
\begin{subfigure}{0.16\linewidth}
  \centering
  \includegraphics[width=\linewidth]{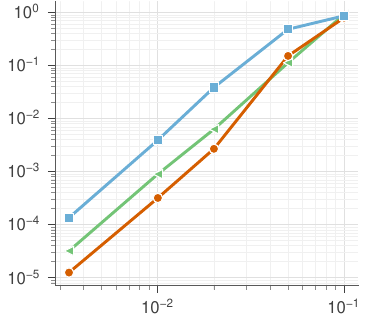}
\end{subfigure}
\begin{subfigure}{0.16\linewidth}
  \centering
  \includegraphics[width=\linewidth]{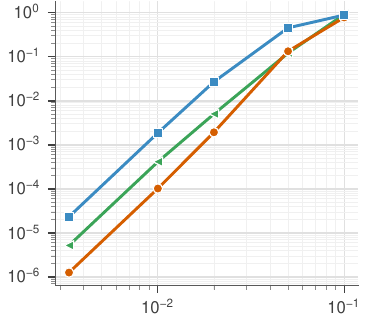}
\end{subfigure}
\begin{subfigure}{0.16\linewidth}
  \centering
  \includegraphics[width=\linewidth]{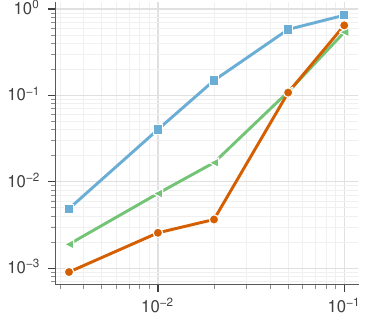}
\end{subfigure}
\begin{subfigure}{0.16\linewidth}
  \centering
  \includegraphics[width=\linewidth]{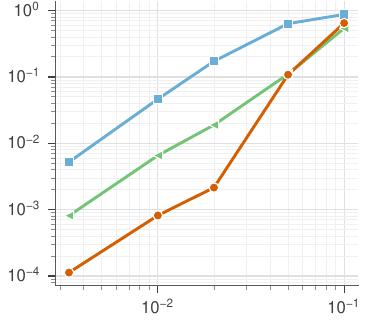}
\end{subfigure}
\begin{subfigure}{0.16\linewidth}
  \centering
  \includegraphics[width=\linewidth]{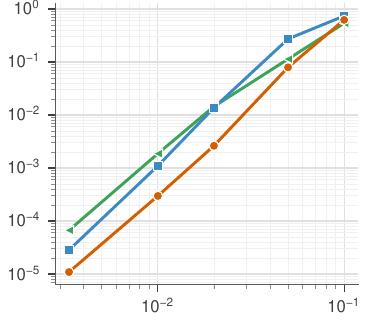}
\end{subfigure}

\end{center}
\caption{Convergence of the error in $\partial_x\phi$ (left three panels)
and $\nabla^2\phi$ (right three panels) for the test function (\eqref{eq:test_function}), at approximation orders $p=2,3,4$ from
left to right within each group. The vertical axis is the relative $L_2$
error, the horizontal axis the average node spacing. SpeND is compared
against the two baselines at matched order and stencil size.}
\label{fig:test_conv}
\end{figure}

Figure~\ref{fig:test_conv} shows the error against mean spacing $s$, with the node count scaling as $s^{-2}$ in two dimensions. At the coarsest resolution, $s = 10^{-1}$, all three methods coincide for every operator and order. The highest harmonic, $k_x = 14\pi$, sits at $k_x s/\pi = 1.4$ of the Nyquist wavenumber $\pi/s$, so no discretisation recovers it, hence the errors are identical. As $s$ decreases the harmonics enter the resolvable band, SpeND separates from both baselines and attains the lowest error at every order.

For $\partial_x$, SpeND reaches a given error with roughly $2\times$ fewer nodes than RBF-FD at every order at fine resolutions. Against LABFM the saving depends strongly on order, falling from roughly $45\times$ at $p = 2$ to $4\times$ at $p = 4$. For $\nabla^2$ the improvement is stronger still, particularly at $p = 2$ and $p = 3$, where, in the best case, SpeND requires up to $15\times$ fewer nodes than RBF-FD to reach the same error. 

\subsection{Viscous Burgers' Equation}
\label{sec:burgers}
We next solve the viscous Burgers equation on the doubly periodic unit square at Reynolds number $100$, with $\mathbf{u} = (u, v)$ and $\mathbf{u}(x,y,0) = (\sin 2\pi x,\, 0)$. The solution remains one-dimensional, the sinusoid steepening into a viscous shock at $x = 0.5$ before decaying, but is discretised with two-dimensional operators. A reference solution is obtained to high precision through the Cole--Hopf transformation. Nodes are placed uniformly and each coordinate is perturbed by $\varepsilon s\,\mathcal{U}(-\tfrac{1}{2}, \tfrac{1}{2})$ with $\varepsilon = 0.8$. This level of disorder is well beyond what particle simulations typically encounter~\citep{Suchde2023}, since particle shifting~\citep{Xu2009ISPH,Lind2012ISPH} keeps the local spacing uniform while the distribution remains unstructured, and serves here as a stress test. We use the $p = 2$ gradient and Laplacian operators from Section~\ref{sec:toy} without retraining or modification, with $\partial_y$ obtained by applying the $\partial_x$ network to the stencil with swapped coordinates. The right-hand side is assembled from the discrete operators of each method and integrated with classical $4^{\text{th}}$ order Runge--Kutta to $t = 1$, and we report the $L_2$ error in $u$ against the reference at $t=1$.
\begin{figure}[h]
  \begin{center}
      \begin{subfigure}{0.35\linewidth}
      \centering
      \includegraphics[width=\linewidth]{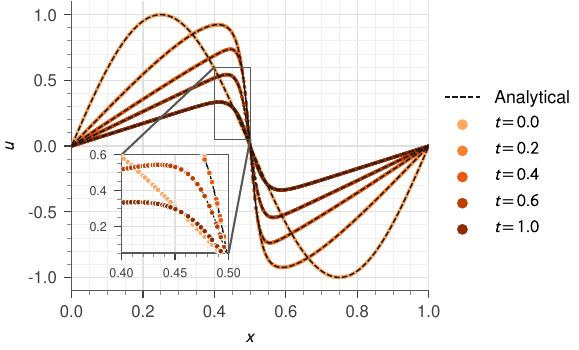}
    \end{subfigure}
    \begin{subfigure}{0.23\linewidth}
      \centering
      \includegraphics[width=\linewidth]{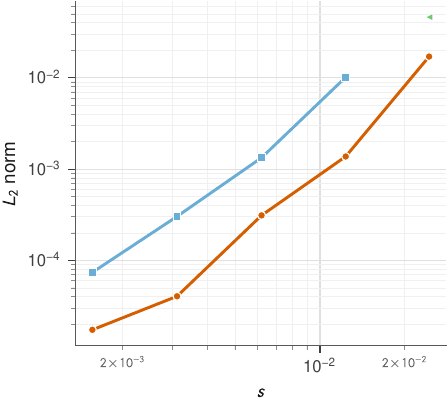}
    \end{subfigure}
    \begin{subfigure}{0.34\linewidth}
      \centering
      \includegraphics[width=\linewidth]{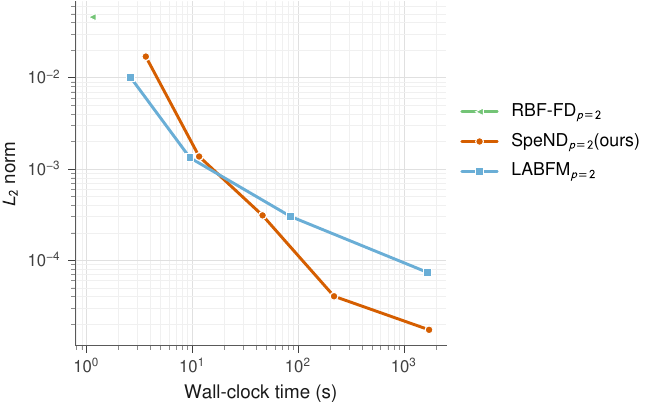}
    \end{subfigure}
  \end{center}
\caption{Viscous Burgers' equation at $\mathit{Re} = 100$. Left, the SpeND solution at successive times against the reference; middle, the $L_2$ error in $u$ at $t = 1$ against average node spacing; right, the same error against total wall-clock time.}
  \label{fig:burgers}
\end{figure}

Figure~\ref{fig:burgers} shows the results. All three methods share the same solver, they differ only in their discrete operators. RBF-FD completes the simulation only at the coarsest resolution and diverges at every finer one. We hypothesise this is due to its dispersion errors (Appendix~\ref{app:mod_resp_train}): the real part of its first-derivative response exceeds the exact response above $0.2k_{\text{Ny}}$, and its Laplacian carries a large imaginary part above $0.4k_{\text{Ny}}$. Dispersion errors of this kind generate spurious oscillations at steep gradients, and as the shock steepens the nonlinear term transfers energy into these wavenumbers, where the oscillations can grow until the solution becomes unstable. LABFM diverges at the coarsest resolution but converges at finer ones. Its modal response departs from the exact one at a higher fraction of $k_{\text{Ny}}$ than SpeND's (Appendix~\ref{app:mod_resp_train}), so at coarse resolution the wavenumbers excited by the steepening shock fall where its errors are large; refining the nodes moves these wavenumbers to a smaller fraction of $k_{\text{Ny}}$, into the range the operator resolves. SpeND completes every run and converges throughout, with an error roughly $10\times$ below LABFM.

Figure~\ref{fig:burgers} right panel shows the error against wall-clock time. Below roughly $3$\,s LABFM attains lower error at equivalent wall-clock time, since at these resolutions SpeND's cost is dominated by its more expensive weight computation. On a fixed node set this one-off cost scales linearly with the number of nodes, whereas time integration scales with the number of nodes times the number of steps. LABFM reaches its finest-resolution error of $7\times10^{-5}$ in about $10^3$\,s, whereas SpeND reaches the same error at a coarser resolution in about $10^2$\,s, a $10\times$ speedup.

\subsection{Poisson's Equation}
\label{sec:poisson}
We solve $\nabla^2\phi = f$ on the unit square with a circular obstacle of radius $0.2$ at its centre, doubly periodic on the outer boundaries and with Dirichlet conditions on the obstacle, using the manufactured solution $\phi = \sin(6\pi x)\sin(6\pi y)$. Nodes are relaxed by particle shifting into the isotropic distribution typical of particle simulations, and stencils adjacent to the obstacle are truncated by the wall. We use the $p = 3$ Laplacian of Section~\ref{sec:toy} without retraining. Since $\nabla^2$ has order $m = 2$, the expected convergence rate is $p + 1 - m = 2$.

\begin{figure}[h]
  \begin{center}
    \begin{subfigure}{0.28\linewidth}
      \centering
      \includegraphics[width=\linewidth]{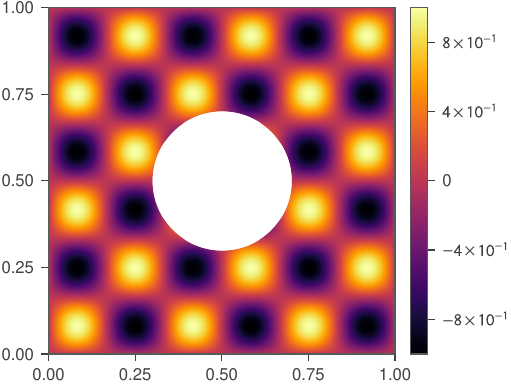}
    \end{subfigure}
    \begin{subfigure}{0.225\linewidth}
      \centering
      \includegraphics[width=\linewidth]{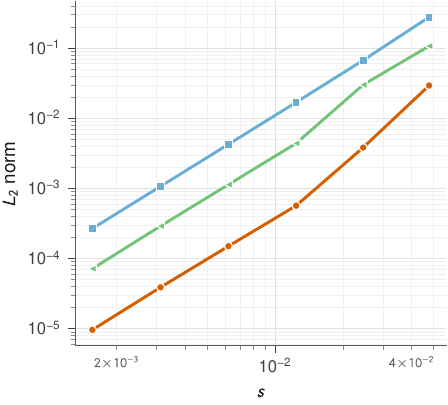}
    \end{subfigure}
    \begin{subfigure}{0.335\linewidth}
      \centering
      \includegraphics[width=\linewidth]{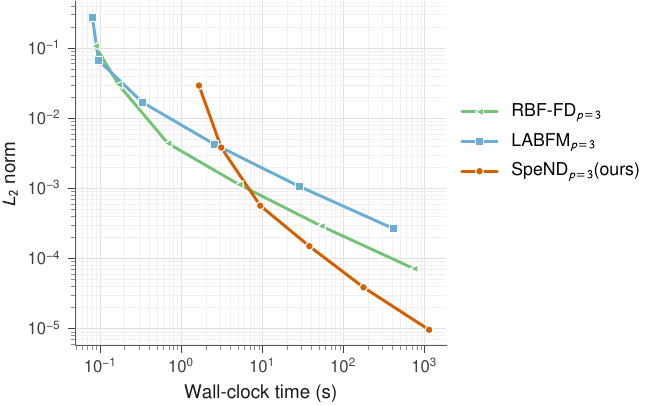}
    \end{subfigure}

  \end{center}
\caption{Poisson equation on the unit square with a central circular obstacle. Left, analytical solution; middle, the $L_2$ error in $\phi$ against average node spacing; right, the $L_2$ error against total wall-clock time.}
  \label{fig:poisson}
\end{figure}

Figure~\ref{fig:poisson} shows the results. All methods converge at second order, and SpeND attains the lowest error at every resolution: at the finest, its error is approximately $28\times$ below LABFM and $7\times$ below RBF-FD. As in Section~\ref{sec:burgers}, SpeND's cost at coarse resolution is dominated by computing the weights, but this is amortised rapidly with refinement. RBF-FD reaches its lowest error, $7\times10^{-5}$, in about $740$\,s, which SpeND matches in about $90$\,s, a speedup of roughly $8\times$. LABFM reaches $3\times10^{-4}$ in about $400$\,s, which SpeND matches in about $20$\,s, roughly
$20\times$ faster.

\subsection{Weakly Compressible Navier--Stokes}
\label{sec:tgv}
Lastly, we simulate the decaying two-dimensional Taylor--Green vortex with the weakly compressible Navier--Stokes equations at Mach number $0.1$ and Reynolds number of $100$ on the periodic domain $[-0.5, 0.5]^2$. Nodes are relaxed by particle shifting as in Section~\ref{sec:poisson}. All methods use $p = 4$ gradient and Laplacian operators, which for SpeND are those of Section~\ref{sec:toy} used without retraining. All runs share the same node sets, time integrator and hyperviscous filter, and differ only in the discrete operators.
\begin{figure}[h]
  \begin{center}
    \begin{subfigure}{0.39\linewidth}
      \centering
      \includegraphics[width=\linewidth]{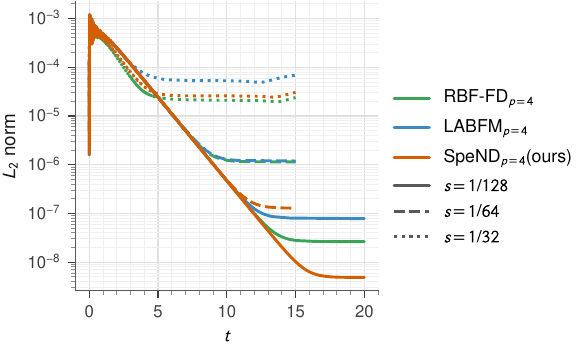}
    \end{subfigure}
    \begin{subfigure}{0.245\linewidth}
      \centering
      \includegraphics[width=\linewidth]{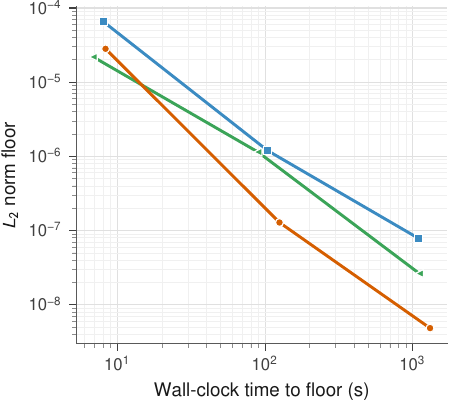}
    \end{subfigure}
  \end{center}
\caption{Taylor--Green vortex at Mach $0.1$ and Reynolds number $100$. Left: $L_2$ error in the velocity, normalised by the velocity magnitude at the first output time, against time at node spacings $s = 1/32$ and $1/64$ (to $t = 15$) and $s = 1/128$ (to $t = 20$). Right: error floor of each run against the wall-clock time taken to reach it, for each method and resolution.}
  \label{fig:tgv}
\end{figure}

At the coarsest resolution ($s = 1/32$) the three schemes plateau at the same order of magnitude, RBF-FD attaining $2.0\times10^{-5}$ and SpeND $2.5\times10^{-5}$. The separation appears with refinement. At $s = 1/64$ SpeND is roughly an order of magnitude below both baselines, and at $s = 1/128$ RBF-FD and LABFM plateau at $2.8\times10^{-8}$ and $8\times10^{-8}$, while SpeND reaches approximately $5\times10^{-9}$.

The advantage persists once cost is accounted for. Plotting the error floor against wall-clock time, all three methods sit close together at about $20$\,s. As resolution increases the curves separate. RBF-FD reaches its lowest error floor, $2.8\times10^{-8}$, in about $10^3$\,s, which SpeND matches in about $3.5\times10^2$\,s, a speedup of roughly $3\times$. LABFM reaches $8\times10^{-8}$ in about $10^3$\,s, which SpeND matches in about $1.8\times10^2$\,s, roughly $5.5\times$ faster.

\section{Conclusion}
We have introduced a framework for learning differential operators on unstructured point clouds, in which a network is trained per operator to optimise the modal response while polynomial consistency is enforced by construction. Formal order of accuracy is therefore preserved, and the training objective carries a direct numerical interpretation rather than acting as a generic regression loss. The same trained operators are applied without retraining to an analytic test function, to Burgers' equation, to a Poisson problem with an embedded obstacle and to the Taylor--Green vortex, reaching matched accuracy between $3\times$ and $20\times$ faster than the classical baselines at the finer resolutions.

In the experiments, at coarse resolutions, the classical baselines remain competitive; the gains appear once the active wavenumbers fall inside the band on which the modal response is optimised. Because the learned operators enter a solver only through their stencil weights, they can replace the operators of an existing mesh-free code without altering the rest of the pipeline, and without giving up the formal order of accuracy that code relies on. More broadly, we see this as a step towards learned components for physical simulation that retain the structural properties classical numerical methods are built on.

\section*{AI use statement}

In this work, we used generative AI tools to help implement methods, primarily providing feedback on the experimental methodology, in particular on how to construct fair comparisons between methods; to help refine the paper's claims in light of the obtained results; and to check and correct the logic of the authors' mathematical derivations. We have not used generative AI tools to develop the conceptual framework, formulate mathematical claims, propose hypotheses, generate synthetic data, or interpret results, and the remaining required disclosure tasks (providing critical ingredients for or writing proofs, language translation, dataset cleaning, qualitative data analysis) are not applicable to this work. 

Additionally, we used generative AI tools to debug, edit, reorganise, and comment on existing software code, to identify relevant literature when studying specific ideas, to search for related work during the literature review, and to edit the paper for readability. We have reviewed all AI-assisted work: all AI-generated code was verified and tested for correctness by the authors, all mathematical content was derived by the authors and any AI-suggested corrections were independently verified, all claims were checked by the authors against the experimental evidence, all AI-suggested references were read and verified against the original sources, and all edited text was reviewed by the authors. We take responsibility for the final content of this work, including text, claims or artifacts produced with the aid of generative AI.

\section*{Ethics statement}

This work develops learned discretisation operators for mesh-free numerical methods and does not involve human subjects, personal data, or sensitive information. All training and evaluation data are synthetic point distributions and analytical test functions generated by the authors, so no dataset licensing, privacy, or consent issues arise. The methods target general-purpose scientific computing, and we do not foresee direct harmful applications beyond those of numerical simulation tools in general.

\section*{Reproducibility statement}

All data in this work were generated by the authors. The training corpus consists only of synthetic stencil geometries, and every test case has an analytical or semi-analytical reference solution, so no external datasets are required. Upon publication, we will release as open source the data generation pipeline, the training code and the trained SpeND operators, together with the Fortran solver used for all experiments, its TorchFort interface to the trained networks, and our implementations of the LABFM and RBF-FD baselines with the configurations used in this paper. Within the paper, the problem setup, moment projection and training objective are described in Sections~\ref{sec:setup}--\ref{sec:training}. Data generation is detailed in Appendix~\ref{app:data}, the architecture and training hyperparameters in Appendix~\ref{app:architecture}, the modal-response objective for other operators in Appendix~\ref{app:resolving}, the baselines in Appendix~\ref{app:operators}, the test-case setups in Appendix~\ref{app:experiments}, and hardware, software and training cost in Appendix~\ref{app:hardware}.


\clearpage

\bibliography{iclr2027_conference}
\bibliographystyle{iclr2027_conference}

\clearpage
\appendix
\section*{Appendix}
\startcontents[appendix]
\printcontents[appendix]{}{1}{\setcounter{tocdepth}{2}}
\clearpage
\section*{Nomenclature}
\phantomsection
\addcontentsline{toc}{section}{Nomenclature}
\label{app:nomenclature}
{\small
\begin{longtable}{@{}p{0.17\linewidth}p{0.79\linewidth}@{}}
\toprule
\multicolumn{2}{@{}l}{\textbf{Domain and discretisation}} \\
\midrule
$\Omega \subset \mathbb{R}^d$ & Spatial domain of dimension $d$ \\
$\mathcal{P}$ & Point cloud (collocation nodes) in $\Omega$ \\
$\mathbf{x}_i$ & Position of node $i$ \\
$(\cdot)_{ji}$ & Difference $(\cdot)_j - (\cdot)_i$ \\
$\mathcal{N}_i$ & Index set of the stencil of node $i$, node $i$ included \\
$N$ & Stencil size, $N = |\mathcal{N}_i|$ \\
$R_i$ & Distance from $\mathbf{x}_i$ to the farthest node of $\mathcal{N}_i$ \\
$s$ & (Mean) node spacing \\
$\phi$ & Generic scalar field \\
\midrule
\multicolumn{2}{@{}l}{\textbf{Operators and consistency}} \\
\midrule
$D$ & Continuous differential operator, e.g.\ $\partial_x$, $\partial_y$, $\nabla^2$ \\
$m$ & Order of $D$ \\
$L^D$ & Discrete approximation of $D$ \\
$w^D_{j,i}$, $\mathbf{w}^D_i$ & Physical stencil weight of node $j$ at node $i$; weight vector of node $i$ \\
$p$ & Consistency order (polynomial degree reproduced exactly) \\
$\boldsymbol{\alpha}$ & Multi-index, $1 \le |\boldsymbol{\alpha}| \le p$ \\
$N_p$ & Number of moment conditions, $\binom{p+d}{d} - 1$ \\
$\mathbf{V}_i$ & Moment matrix of node $i$ \\
$\mathbf{d}^D$ & Stencil moment vector of $D$ \\
\midrule
\multicolumn{2}{@{}l}{\textbf{SpeND}} \\
\midrule
$\bar{(\cdot)}$ & Quantity in normalised stencil coordinates \\
$\bar{\mathbf{x}}_{ji}$ & Normalised offset, $\mathbf{x}_{ji}/R_i$ \\
$\bar{\mathbf{V}}_i$ & Moment matrix assembled from $\bar{\mathbf{x}}_{ji}$ \\
$f_\theta$, $\theta$ & Weight-predicting network and its parameters \\
$\tilde{\mathbf{w}}^D_i$ & Candidate (unprojected) weights \\
$\Pi_i$ & Moment projection onto the consistent affine subspace \\
$\bar{\mathbf{w}}^D_i$ & Projected weights, $\mathbf{w}^D_i = R_i^{-m}\bar{\mathbf{w}}^D_i$ \\
$\mathbf{P}_i$ & Orthogonal projector onto $\ker \bar{\mathbf{V}}_i$ \\
$(\cdot)^{+}$ & Moore--Penrose pseudoinverse \\
$\sigma_m$ & Output scaling factor, $\bar{s}_N^{-m}$ \\
\midrule
\multicolumn{2}{@{}l}{\textbf{Modal response}} \\
\midrule
$\mathbf{k} = (k_x, k_y)$ & Wavevector \\
$\phi_{\mathbf{k}}$ & Fourier mode $e^{\mathrm{i}\mathbf{k}\cdot\mathbf{x}}$ \\
$\hat{L}^D_i(\mathbf{k})$ & Modal response of $L^D$ at node $i$ \\
$k_\text{eff}$ & Effective wavenumber of $\partial_x$, $-\mathrm{i}\hat{L}^{\partial_x}_i$ \\
$q^2_\text{eff}$ & Effective squared wavenumber of $\nabla^2$, $-\hat{L}^{\nabla^2}_i$ \\
$\bar{s}_N$ & Mean normalised spacing, $\sqrt{\pi/N}$ \\
$k_{\mathrm{Ny}}$ & Nyquist wavenumber, $\pi/\bar{s}_N$ \\
$\hat{k}$ & Normalised wavenumber, $\|\mathbf{k}\|/k_{\mathrm{Ny}}$ \\
$\eta$ & Band limit, $\|\mathbf{k}\| \le \eta\,k_{\mathrm{Ny}}$ \\
\midrule
\multicolumn{2}{@{}l}{\textbf{Training}} \\
\midrule
$\Phi$ & Class of test functions \\
$r_i[\phi]$ & Stencil residual \\
$\mathcal{K}$ & Training wavenumbers on the resolved half-disk \\
$\sigma(\mathbf{k})$ & Relative-error weight \\
$\epsilon$ & Floor of $\sigma(\mathbf{k})$ \\
$\omega_>$, $\lambda_>$ & Asymmetric weight and penalty for over-predicted real part \\
$\omega_+$, $\lambda_+$ & Asymmetric weight and penalty for positive imaginary part \\
$\mathcal{L}^{\text{disp}}_i$, $\mathcal{L}^{\text{diss}}_i$ & Per-stencil dispersion and dissipation losses \\
$\gamma$ & Weight of $\mathcal{L}^{\text{diss}}_i$ \\
$\mathcal{S}$ & Corpus of training stencils \\
$\lambda_{\mathrm{b}}$ & Adaptive boundary-stencil weight \\
$B$, $C$ & Number of network blocks; feature width \\
\midrule
\multicolumn{2}{@{}l}{\textbf{Data generation}} \\
\midrule
$\varepsilon$ & Lattice disorder amplitude \\
$n_{\mathrm{it}}$ & Number of particle-shifting iterations \\
$h$ & Kernel smoothing length (particle shifting, LABFM) \\
$\mathbf{n}$, $\vartheta$ & Wall normal and its angle \\
$\beta$ & Wall distance from the central node, in units of $R_i$ \\
\midrule
\multicolumn{2}{@{}l}{\textbf{Model problems and analysis}} \\
\midrule
$t$ & Time \\
$\mathbf{u} = (u, v)$ & Velocity \\
$\rho$ & Density \\
$f$ & Poisson source term \\
$\mathit{Re}$, $\mathit{Ma}$ & Reynolds and Mach numbers \\
$a$, $\tilde{a}$ & Advection speed; numerical phase speed \\
$\nu$ & Diffusivity \\
$\hat{u}(t)$ & Amplitude of a single Fourier mode \\
$\mathbf{G}^D$, $\mu$ & Global operator matrix and its eigenvalues \\
$\boldsymbol{\phi}$ & Vector of nodal values \\
\midrule
\multicolumn{2}{@{}l}{\textbf{Baselines}} \\
\midrule
$\mathbf{W}_{ji}$, $\boldsymbol{\Psi}^D_i$ & LABFM anisotropic basis functions and coefficients \\
$\mathbf{A}_i$, $\mathbf{X}_{ji}$ & LABFM system matrix; vector of Taylor monomials \\
$\kappa$, $H_n$ & Wendland $C^2$ kernel; Hermite polynomial of degree $n$ \\
$\chi$ & Polyharmonic spline $r^5$ \\
$\mathbf{K}_i$, $\mathbf{M}_i$ & RBF-FD kernel and polynomial matrices \\
$\boldsymbol{\xi}_i$, $\mathbf{c}^D_i$ & RBF-FD Lagrange multipliers; kernel right-hand side \\
\bottomrule
\end{longtable}
}

\clearpage

\clearpage
\section{Data generation}
\label{app:data}
The dataset contains stencil geometries only, with no function values or reference solutions. Since all experiments are two-dimensional, each sample is a two-dimensional stencil of $N$ nodes, the central node included. The generation procedure below extends directly to three dimensions, the only change being the dimension of the node coordinates. Rather than being extracted from a single discretised domain, each stencil is generated from its own independent point cloud. Stencils are stored as the normalised offsets $\bar{\mathbf{x}}_{ji} = \mathbf{x}_{ji}/R_i$, which places every stencil within the unit circle.

\begin{figure}[h]
  \begin{center}
    \begin{subfigure}{0.8\linewidth}
      \centering
      \includegraphics[width=\linewidth]{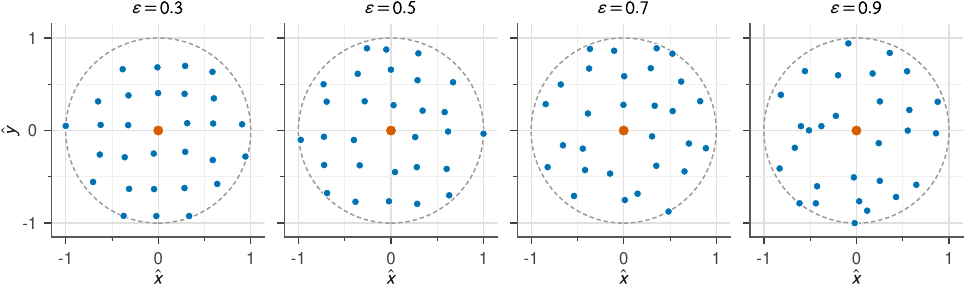}
    \end{subfigure}
    
    \begin{subfigure}{0.8\linewidth}
      \centering
      \includegraphics[width=\linewidth]{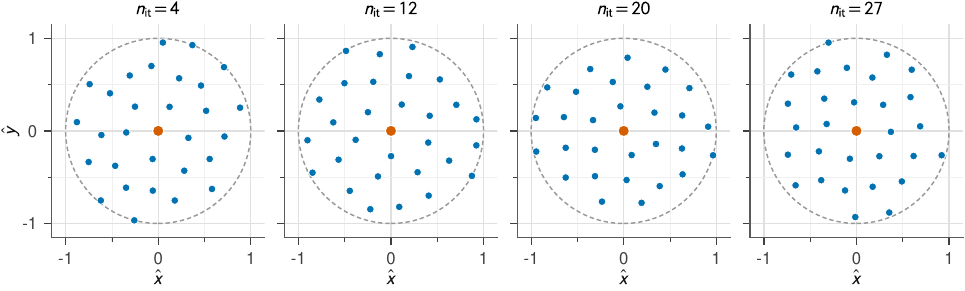}
    \end{subfigure}
  \end{center}
\caption{Representative stencils from each of the eight bins. Top row, perturbed Cartesian lattices with increasing disorder $\varepsilon$. Bottom row, particle-shifted node sets with increasing number of shifting iterations $n_{\mathrm{it}}$. The value of $\varepsilon$ or $n_{\mathrm{it}}$ is given above each panel.}
  \label{fig:nodes}
\end{figure}

\paragraph{Node distributions.}
Stencils are generated from two families of node sets. The first is a Cartesian lattice in which every node is   displaced independently along each axis by $\varepsilon s\,\mathcal{U}(-\tfrac{1}{2}, \tfrac{1}{2})$, with the disorder $\varepsilon$ drawn uniformly per stencil; this configuration is commonly referred to as random node placement~\citep{Suchde2023}. The second imitates the node sets on which mesh-free simulations are run in practice, and is obtained by iterative particle shifting~\citep{Xu2009ISPH,Lind2012ISPH,FLYER201639}. Starting from a lattice with the high disorder $\varepsilon = 1$, each of $n_{\mathrm{it}}$ iterations pushes every node away from its neighbours within a radius $2h$, driving the cloud towards a near-isotropic packing. We use $h = 2s$ and cap each displacement at $0.2s$, so that $n_{\mathrm{it}} = 0$ yields the disordered lattice and $n_{\mathrm{it}} = 30$ gives a relaxed packing. Each family is divided into four bins of equal size, $\varepsilon \in [0.2, 0.4], [0.4, 0.6], [0.6, 0.8], [0.8, 1.0]$ and $n_{\mathrm{it}} \in \{0,\dots,7\}, \{8,\dots,15\}, \{16,\dots,23\}, \{24,\dots,30\}$, with the parameter drawn uniformly within each bin. A representative stencil from each bin is shown in Figure~\ref{fig:nodes}.

\begin{figure}[h]
  \begin{center}
    \begin{subfigure}{0.8\linewidth}
      \centering
      \includegraphics[width=\linewidth]{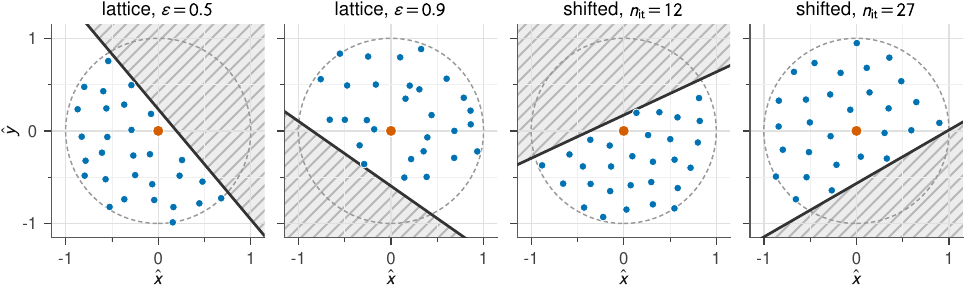}
    \end{subfigure}

  \end{center}
\caption{Representative boundary stencils. The two left panels are drawn from perturbed Cartesian lattices and the two right panels from particle-shifted node sets. The bold line marks the wall.}
  \label{fig:nodes_bound}
\end{figure}

\paragraph{Boundary stencils.}
Every interior stencil is paired with one copy that imitates a stencil adjacent to a wall. We draw a line with unit normal $\mathbf{n} = (\cos\vartheta, \sin\vartheta)$, $\vartheta \sim \mathcal{U}(0, 2\pi)$, at a distance $\beta \sim \mathcal{U}(0.05, 0.7)$ from the central node in units of $R_i$, and discard every node with $\bar{\mathbf{x}}_{ji} \cdot \mathbf{n} > \beta$. The stencil is then refilled to $N$ nodes with the nearest remaining nodes on the domain side and renormalised by its new $R_i$. Since $\beta > 0$ the central node always remains in the domain, and every boundary stencil keeps exactly $N$ nodes. Representative boundary stencils from both families are shown in Figure~\ref{fig:nodes_bound}.

\paragraph{Dataset size and split.}
Each of the eight bins contributes $5 \times 10^5$ interior stencils and their $5 \times 10^5$ boundary copies, giving $8 \times 10^6$ stencils in total, half of them boundary stencils. These are split at random into training, validation and test sets in the proportions $70/20/10$.

\section{Architecture \& Training Details}
\label{app:architecture}

\paragraph{Architecture.}
Each node $j \in \mathcal{N}_i$ enters the network only through its normalised offset $\bar{\mathbf{x}}_{ji}$. A shared two-layer encoder of widths $128$ and $C = 256$ lifts every offset to $\mathbf{h}_j^{(0)} \in \mathbb{R}^{C}$. The features are then updated by $B = 4$ blocks with independent parameters,
\begin{equation}
\mathbf{z}_j = \tanh\big(W_1^{(\ell)}\mathbf{h}_j^{(\ell-1)} + \mathbf{b}_1^{(\ell)}\big), \qquad
\mathbf{g} = \max_{j' \in \mathcal{N}_i} \mathbf{z}_{j'}, \qquad
\mathbf{h}_j^{(\ell)} = \tanh\big(W_2^{(\ell)}[\mathbf{z}_j ; \mathbf{g}] + \mathbf{b}_2^{(\ell)}\big),
\end{equation}
where the maximum is taken elementwise. Pooling occurs in every block, so each node is conditioned on the whole stencil repeatedly while the output remains one value per node. A shared decoder of widths $128$ and $1$, with a linear final layer, returns the candidate weight of each node. All maps are shared across nodes and the pooling is symmetric. The weights are therefore permutation equivariant, and no parameter depends on $N$. The network uses $\tanh$ activations for a total of roughly $855$ thousand parameters. A separate network with this architecture is trained for each discrete differential operator.

\begin{figure}[h]
  \begin{center}
    \begin{subfigure}{1.0\linewidth}
      \centering
      \includegraphics[width=\linewidth]{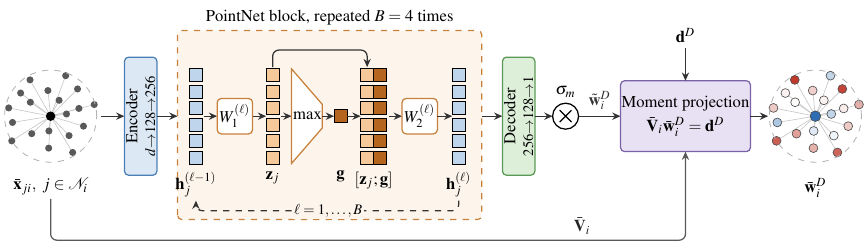}
    \end{subfigure}

  \end{center}
\caption{Architecture of SpeND. Normalised offsets $\bar{\mathbf{x}}_{ji}$ are encoded, refined by $B = 4$ PointNet blocks with elementwise max pooling over the stencil, and decoded into unprojected weights $\tilde{\mathbf{w}}_i^D$. The moment projection enforces $\bar{\mathbf{V}}_i \bar{\mathbf{w}}_i^D = \mathbf{d}^D$ exactly.}
  \label{fig:architecture}
\end{figure}

\paragraph{Output scaling.}
The weights of an operator of order $m$ scale as the node spacing to the power $-m$. The decoder output is therefore multiplied by the fixed factor $\sigma_m = \bar{s}_N^{-m}$ before projection, where $\bar{s}_N = \sqrt{\pi/N}$ is the mean spacing of $N$ nodes filling the unit disk. Without this factor, the decoder would need to produce weight norms of varying orders of magnitude. With $\sigma_m$, the weights are normalised independently of the order of the operator. 

\paragraph{Training.}
A base network is trained on stencils of $N = 25$ for $40$ epochs with Adam at learning rate $3\times10^{-4}$, and then $20$ epochs with Muon \citep{jordan2024muon}. Operators for each configuration, differing in stencil size, consistency order or boundary treatment, are obtained by fine-tuning this base network with Muon for $40$ epochs, with a narrower training band and a smaller $\epsilon$. Fine-tuning from a common base reduces training cost rather than improving the final operators. Muon applies only to weight matrices, so during these phases the biases are updated with Adam. The learning rates are $2.7\times10^{-3}$ for Muon and $9\times10^{-5}$ for Adam, with a linear warm-up over $1000$ optimiser steps. All phases use cosine decay to $10^{-7}$ and batches of $1024$ stencils. The loss on boundary stencils is weighted by $\lambda_{\mathrm{b}}$, set each epoch to the square root of the boundary-to-interior loss ratio. Remaining hyperparameters are listed in Table~\ref{tab:hyperparameters}.

\begin{table}[h]
\centering
\caption{Training hyperparameters of the modal-response objective for the base network and the fine-tuned operators. Operators with $p = 2$ and $p = 4$ are trained with these values. Operators with $p = 3$ are not trained separately, and instead reuse the $p = 2$ network with the $p = 3$ moment constraints imposed by the projection.}
\label{tab:hyperparameters}
\begin{tabular}{lcccc}
\toprule
 & \multicolumn{2}{c}{Base} & \multicolumn{2}{c}{Fine-tuned} \\
\cmidrule(lr){2-3}\cmidrule(lr){4-5}
 & $\partial_x$ & $\nabla^2$ & $\partial_x$ & $\nabla^2$ \\
\midrule
Band limit $\eta$ & $0.7$ & $0.7$ & $0.4$ & $0.4$ \\
$\epsilon$ & $0.25$ & $0.1$ & $0.1$ & $0.1$ \\
Over-prediction penalty, real part $\lambda_>$ & $50$ & $50$ & $50$ & $50$ \\
Over-prediction penalty, imaginary part $\lambda_+$ & $10$ & $1$ & $10$ & $1$ \\
Imaginary-part weight $\gamma$ & $1$ & $1$ & $1$ & $1$ \\
Boundary weight $\lambda_{\mathrm{b}}$ & \multicolumn{4}{c}{adaptive, clipped to $[1, 50]$} \\
\bottomrule
\end{tabular}
\end{table}

\section{Modal Response \& Analysis}
\label{app:resolving}
\subsection{Modal Response of the Laplacian}
\label{app:resolving_operators}
In Section~\ref{sec:training} we illustrate how the exact modal response is obtained for $\partial_x$. Here we repeat the procedure for the Laplacian, which illustrates how it extends to other operators.

As before, we place $\mathbf{x}_i$ at the origin and apply the continuous operator to a Fourier mode, which gives $\nabla^2 e^{\mathrm{i}\mathbf{k}\cdot\mathbf{x}}\big|_{\mathbf{x}=\mathbf{0}} = -\|\mathbf{k}\|^2$. The exact response is now real and negative, so we normalise the discrete response by $-1$ and define the effective squared wavenumber $q^2_\text{eff}(\mathbf{k}) := -\hat{L}^{\nabla^2}_i(\mathbf{k})$, which the exact operator reproduces as $\|\mathbf{k}\|^2$~\citep{Lele1992}. Separating real and imaginary parts,
\begin{equation}
\label{eq:keff_lap}
    q^2_\text{eff}(\mathbf{k}) = \sum_{j \in \mathcal{N}_i} \bar{w}_{j,i}^{\nabla^2}
    \left[ 1-\cos(\mathbf{k} \cdot \bar{\mathbf{x}}_{ji}) \right]
    \;-\; \mathrm{i} \sum_{j \in \mathcal{N}_i} \bar{w}_{j,i}^{\nabla^2}
    \sin(\mathbf{k} \cdot \bar{\mathbf{x}}_{ji}).
\end{equation}
Compared with \eqref{eq:keff}, the two parts exchange roles. The real part now carries the physical response and is compared against $\|\mathbf{k}\|^2$, while the imaginary part has no continuous counterpart and should vanish. The parity also reverses, since the real part is now even in $\mathbf{k}$ and the imaginary part odd, so the half-disk $\mathcal{K}$ again suffices.

The losses of Section~\ref{sec:training} carry over with $k_\text{eff}$ replaced by $q^2_\text{eff}$ and $k_x$ by $\|\mathbf{k}\|^2$ in \eqref{eq:loss_disp}, so that $\lambda_>$ now penalises a real part exceeding the exact response, which corresponds to excess diffusion, and $\sigma(\mathbf{k}) = \max(\|\mathbf{k}\|^2/k_{\mathrm{Ny}}^2, \epsilon)^{-2}$. Since the roles of the two parts are exchanged, \eqref{eq:loss_disp} now controls dissipation and \eqref{eq:loss_diss} dispersion. The imaginary part is penalised symmetrically, with $\lambda_+ = 1$. Unlike for $\partial_x$, where a leading response exceeds a nonzero physical speed, the exact drift here is zero, so both signs produce a spurious drift of equal speed in opposite directions and neither is the faster side (Appendix~\ref{app:role}).

Other operators follow the same procedure. Evaluating the continuous operator on $e^{\mathrm{i}\mathbf{k}\cdot\mathbf{x}}$ at the origin gives a real polynomial in $\mathbf{k}$ multiplied by $\mathrm{i}^m$, where $m$ is the order of the operator. Dividing the discrete response by $\mathrm{i}^m$ makes the exact response real, and the loss then compares the part that carries it while driving the other part to zero.

\subsection{Modal Response of the Trained Operators}
\label{app:mod_resp_train}
We report the modal response of the operators used in Section~\ref{sec:experiments}, comparing SpeND against the classical baselines for each operator and consistency order. The modal response measures how a discrete operator acts on a single Fourier mode, compared with the exact operator, with the real and imaginary parts governing dispersion and dissipation respectively (Section~\ref{sec:training}).

\begin{figure}[h]
  \begin{center}
    \centering
\includegraphics[width=0.5\linewidth]{manuscript/test_convergence/legend_horizontal.pdf}\\

    \begin{subfigure}{0.32\linewidth}
      \centering
      \includegraphics[width=\linewidth]{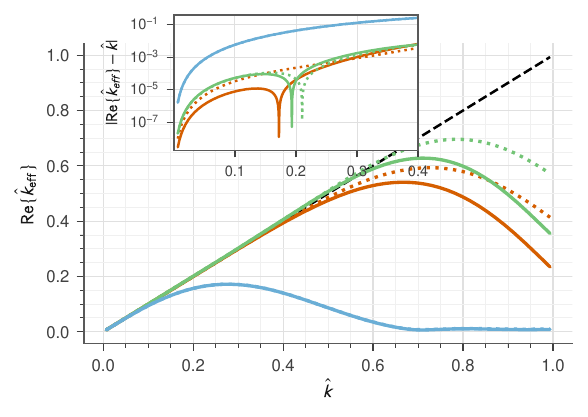}
    \end{subfigure}
    \begin{subfigure}{0.32\linewidth}
      \centering
      \includegraphics[width=\linewidth]{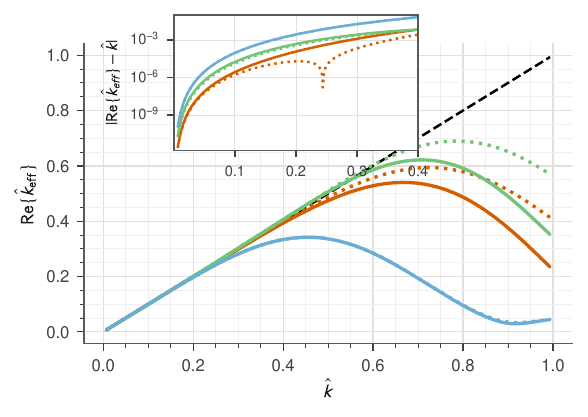}
    \end{subfigure}
        \begin{subfigure}{0.32\linewidth}
      \centering
      \includegraphics[width=\linewidth]{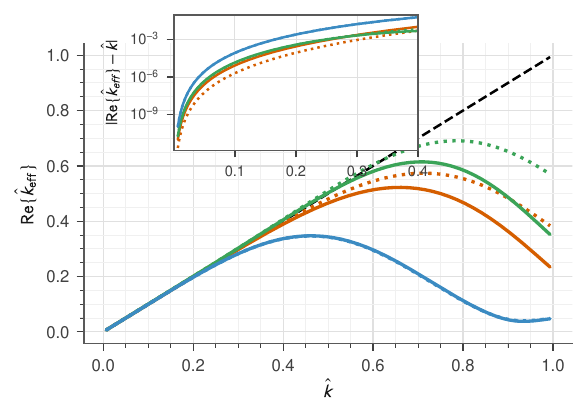}
    \end{subfigure}
    
    \begin{subfigure}{0.32\linewidth}
      \centering
      \includegraphics[width=\linewidth]{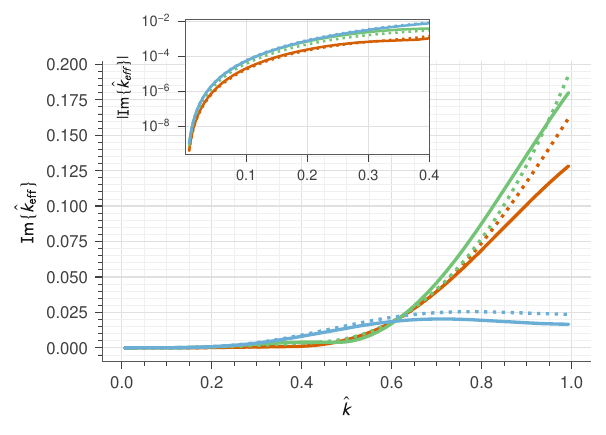}
    \end{subfigure}
    \begin{subfigure}{0.32\linewidth}
      \centering
      \includegraphics[width=\linewidth]{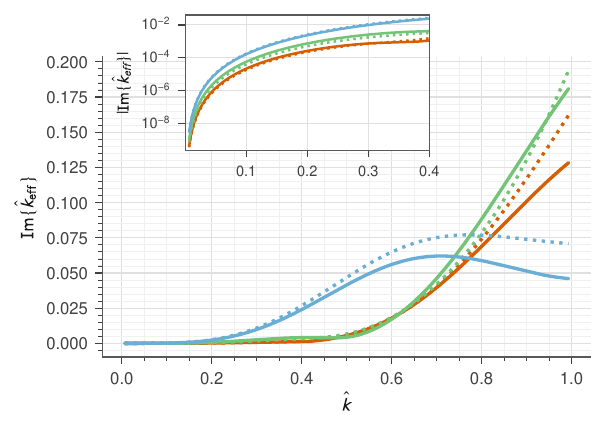}
    \end{subfigure}
    \begin{subfigure}{0.32\linewidth}
      \centering
      \includegraphics[width=\linewidth]{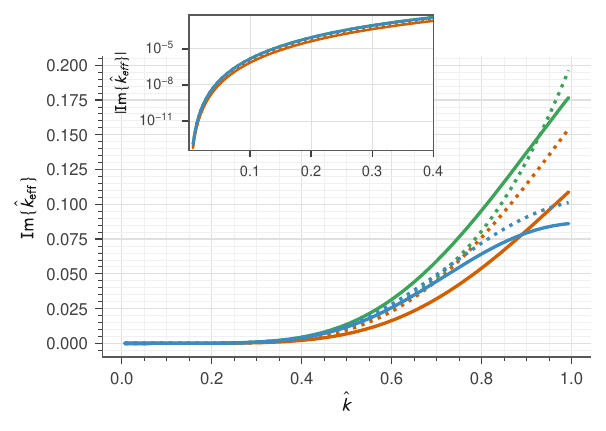}
    \end{subfigure}
    
    \begin{subfigure}{0.32\linewidth}
      \centering
      \includegraphics[width=\linewidth]{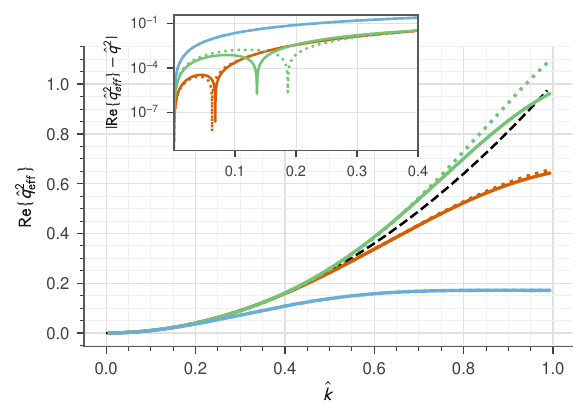}
    \end{subfigure}
        \begin{subfigure}{0.32\linewidth}
      \centering
      \includegraphics[width=\linewidth]{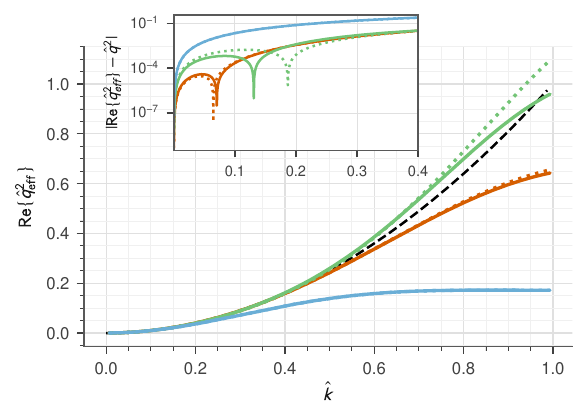}
    \end{subfigure}
        \begin{subfigure}{0.32\linewidth}
      \centering
      \includegraphics[width=\linewidth]{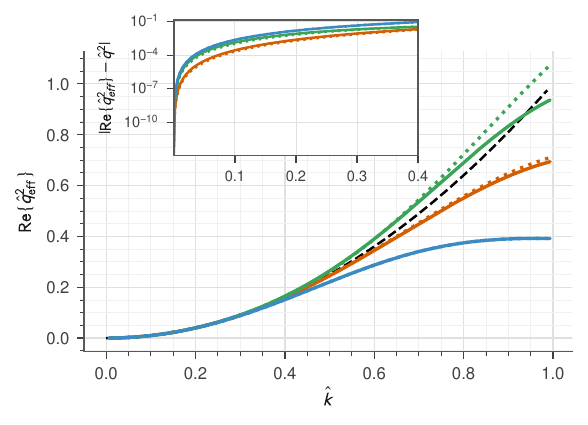}
    \end{subfigure}
    
    \begin{subfigure}{0.32\linewidth}
      \centering
      \includegraphics[width=\linewidth]{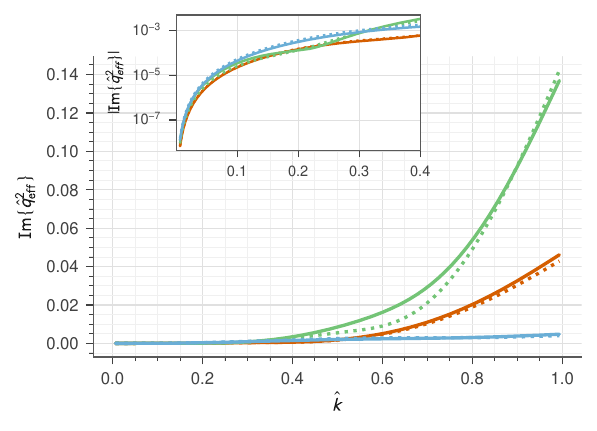}
    \end{subfigure}
    \begin{subfigure}{0.32\linewidth}
      \centering
      \includegraphics[width=\linewidth]{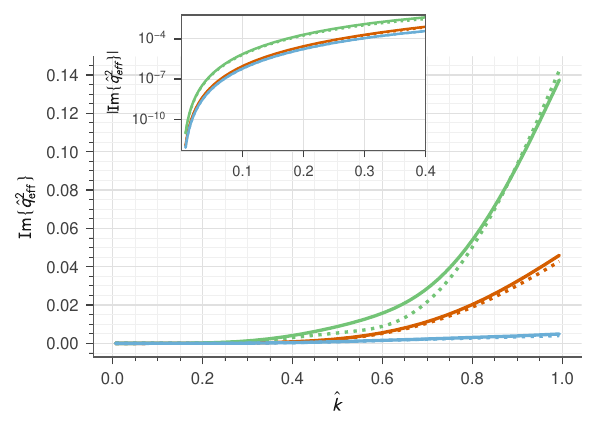}
    \end{subfigure}
    \begin{subfigure}{0.32\linewidth}
      \centering
      \includegraphics[width=\linewidth]{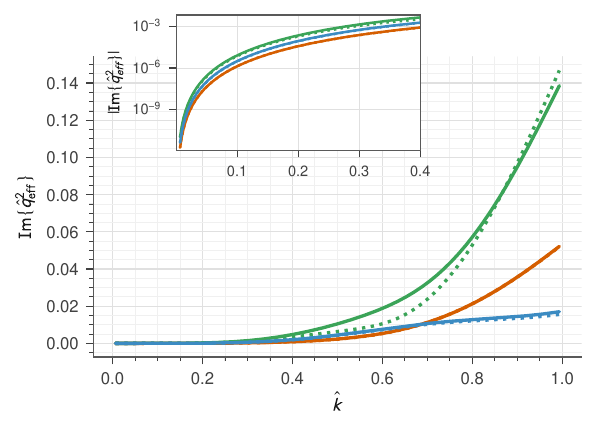}
    \end{subfigure}

  \end{center}
\caption{Modal response of the $\partial_x$ (rows one and two) and $\nabla^2$ (rows three and four) operators, showing the real and imaginary parts for consistency orders $p = 2, 3, 4$ (left to right). Wavenumbers are normalised by $k_{\mathrm{Ny}}$. Solid lines correspond to modes aligned with the $x$-axis, $k_y = 0$, and dotted lines to diagonal modes, $k_x = k_y$. The dashed line is the exact response, and the insets show the error on a logarithmic scale. Stencils were generated with iterative particle shifting with $N=30$.}
    \label{fig:modal}
\end{figure}

Figure~\ref{fig:modal} shows the real and imaginary parts of the modal response of the $\partial_x$ and $\nabla^2$ operators for consistency orders $p = 2, 3, 4$ against wavenumbers normalised by the Nyquist wavenumber. The operators were trained with $\eta = 0.4$, so the objective constrains the modal response only for $\hat{k} \le 0.4$. Within this band, SpeND matches or improves on both baselines in every panel, in both the real and imaginary parts. The largest gains occur at the lowest wavenumbers, where the logarithmic insets show errors up to several orders of magnitude below LABFM and consistently below RBF-FD. Beyond the training band the response is unconstrained, and the ranking between methods varies with operator and order.

\subsection{Dispersion and Dissipation in Advection and Diffusion}
\label{app:role}
We show how the two parts of the modal response enter the solution of the advection and diffusion equations. In both cases we discretise in space only, keeping time continuous, and substitute a single Fourier mode $u(\mathbf{x},t) = \hat{u}(t)\,e^{\mathrm{i}\mathbf{k}\cdot\mathbf{x}}$.

\paragraph{Advection.} For the one-dimensional equation $\partial_t u + a\,\partial_x u = 0$ with constant velocity $a > 0$, substituting the Fourier mode gives $\partial_x u = \mathrm{i}k_x u$, so $\hat{u}(t) = \hat{u}(0)\,e^{-\mathrm{i}a k_x t}$ and $u = \hat{u}(0)\,e^{\mathrm{i}k_x(x - at)}$, a wave travelling at speed $a$. Replacing $\partial_x$ by its discrete counterpart, whose response we write as $\hat{L}^{\partial_x}_i = \mathrm{i}\,k_\text{eff}$ (so that $k_\text{eff} = k_x$ for an exact operator), gives $\mathrm{d}\hat{u}/\mathrm{d}t = -\mathrm{i}a\,k_\text{eff}\,\hat{u}$. Splitting $k_\text{eff}$ into real and imaginary parts, its solution is
\begin{equation}
\label{eq:role_adv}
    \hat{u}(t) = \hat{u}(0)\,
    \underbrace{e^{-\mathrm{i}a\operatorname{Re}\{k_\text{eff}\}\,t}}_{\text{phase}}\,
    \underbrace{e^{a\operatorname{Im}\{k_\text{eff}\}\,t}}_{\text{amplitude}}.
\end{equation}
The first factor has a purely imaginary exponent and hence unit modulus, so it rotates $\hat{u}$ in the complex plane without changing its size; the second is real and positive, so it changes the size only. Thus $|\hat{u}(t)| = |\hat{u}(0)|\,e^{a\operatorname{Im}\{k_\text{eff}\}t}$, while multiplying by $e^{\mathrm{i}k_x x}$ turns the phase factor into $e^{\mathrm{i}k_x(x - \tilde{a}t)}$ with $\tilde{a} = a\operatorname{Re}\{k_\text{eff}\}/k_x$, i.e.\ a translation at speed $\tilde{a}$ instead of $a$. Any deviation of $\operatorname{Re}\{k_\text{eff}\}$ from $k_x$ is therefore a dispersion error, with the mode leading when $\operatorname{Re}\{k_\text{eff}\}/k_x > 1$ and lagging if $\operatorname{Re}\{k_\text{eff}\}/k_x < 1$. Any non-zero $\operatorname{Im}\{k_\text{eff}\}$ is a dissipation error: the mode is damped when $\operatorname{Im}\{k_\text{eff}\} < 0$ and amplified when $\operatorname{Im}\{k_\text{eff}\} > 0$.

\paragraph{Diffusion.} For $\partial_t u = \nu\nabla^2 u$ with $\nu > 0$, the exact solution is $\hat{u}(t) = \hat{u}(0)\,e^{-\nu\|\mathbf{k}\|^2 t}$, a stationary mode decaying at rate $\nu\|\mathbf{k}\|^2$. With $\hat{L}^{\nabla^2}_i = -q^2_\text{eff}$, the semi-discrete solution is
\begin{equation}
\label{eq:role_diff}
    \hat{u}(t) = \hat{u}(0)\,
    \underbrace{e^{-\nu\operatorname{Re}\{q^2_\text{eff}\}\,t}}_{\text{amplitude}}\,
    \underbrace{e^{-\mathrm{i}\nu\operatorname{Im}\{q^2_\text{eff}\}\,t}}_{\text{phase}}.
\end{equation}
In this case, the real part sets the decay rate, so $\operatorname{Re}\{q^2_\text{eff}\} > \|\mathbf{k}\|^2$ over-damps the mode, a smaller positive value under-damps it. The imaginary part sets a phase that is absent from the exact solution, translating the mode at speed $\nu\operatorname{Im}\{q^2_\text{eff}\}/\|\mathbf{k}\|$.

\section{Stability Analysis}
\label{app:stability}
The modal analysis of Appendix~\ref{app:role} treats each stencil as if it were repeated at every node, i.e.\ as a translation-invariant discretisation, in which case the modal response fully determines stability. On an irregular point cloud neighbouring stencils differ, and since $w^D_{j,i}$ is not tied to $w^D_{i,j}$, the symmetries that constrain the spectrum of a translation-invariant discretisation are not valid. Stencils that are individually well behaved can then still produce growing global modes. Since SpeND is trained on a local objective, its global stability must be verified directly.

We assemble the local operators into a global matrix $\mathbf{G}^D \in \mathbb{R}^{|\mathcal{P}| \times |\mathcal{P}|}$, so that applying $L^D$ at every node is the product $\mathbf{G}^D\boldsymbol{\phi}$, with $\boldsymbol{\phi}$ the vector of nodal values. Under $\mathrm{d}\boldsymbol{\phi}/\mathrm{d}t = \mathbf{G}^D\boldsymbol{\phi}$, global modes with eigenvalue $\mu$ grow if $\operatorname{Re}\{\mu\} > 0$, decay if $\operatorname{Re}\{\mu\} < 0$, and oscillate with constant amplitude if $\mu$ is purely imaginary. The first derivative only translates modes, so an ideal discretisation has $\operatorname{Re}\{\mu\} = 0$ for all $\mu$, whereas the Laplacian only damps them, so a stable discretisation requires $\operatorname{Re}\{\mu\} \le 0$~\citep{Fornberg2011}.

\begin{figure}[h]
  \begin{center}
    \centering
\includegraphics[width=0.5\linewidth]{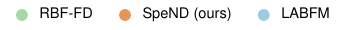}\\

    \begin{subfigure}{0.45\linewidth}
      \centering
      \includegraphics[width=\linewidth]{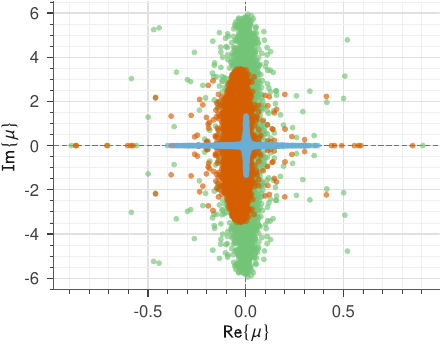}
    \end{subfigure}
    \begin{subfigure}{0.45\linewidth}
      \centering
      \includegraphics[width=\linewidth]{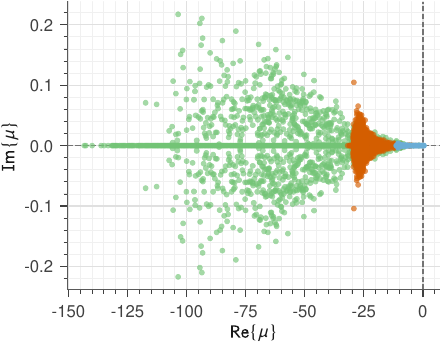}
    \end{subfigure}
    
    \begin{subfigure}{0.45\linewidth}
      \centering
      \includegraphics[width=\linewidth]{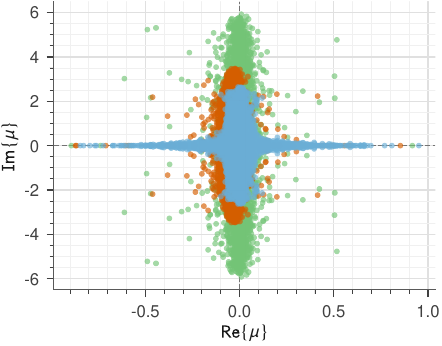}
    \end{subfigure}
    \begin{subfigure}{0.45\linewidth}
      \centering
      \includegraphics[width=\linewidth]{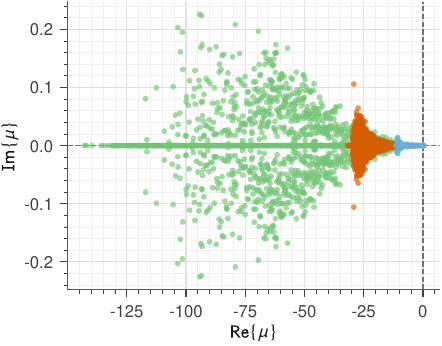}
    \end{subfigure}

    \begin{subfigure}{0.45\linewidth}
      \centering
      \includegraphics[width=\linewidth]{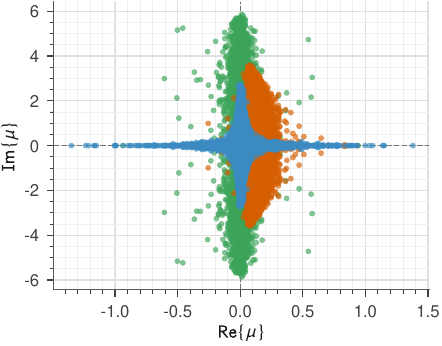}
    \end{subfigure}
        \begin{subfigure}{0.45\linewidth}
      \centering
      \includegraphics[width=\linewidth]{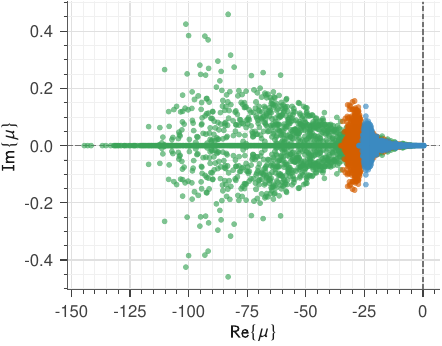}
    \end{subfigure}

  \end{center}
\caption{Normalised eigenvalues $\mu$ of the global operator $\mathbf{G}^D$ on a node set generated by iterative particle shifting, for $\partial_x$ (left) and $\nabla^2$ (right) at consistency orders $p = 2, 3, 4$ (top to bottom). The ideal spectrum lies on the imaginary axis for $\partial_x$ and on the non-positive real axis for $\nabla^2$.}
    \label{fig:stability}
\end{figure}

Figure~\ref{fig:stability} shows the spectra of $\mathbf{G}^D$ on a node set generated by iterative particle shifting, for consistency orders $p = 2, 3, 4$. For both operators, the eigenvalues of RBF-FD generally spread furthest along both axes, followed by SpeND, while those of LABFM are the most concentrated. The one exception is the fourth-order $\partial_x$ operator, for which LABFM spreads furthest along the real axis. Spread away from the ideal axis, along the real axis for $\partial_x$ and the imaginary axis for $\nabla^2$, measures spurious growth, decay or drift, and its ordering is consistent with the imaginary parts of the modal response (Figure~\ref{fig:modal}). Spread along the ideal axis instead reflects the range of wavenumbers the operator resolves, so the compact spectrum of LABFM there, particularly at second order, mirrors the early collapse of its modal response.

\section{Traditional Mesh-Free Discretisations}
\label{app:operators}

Both baselines share the stencil definition of SpeND, with $\mathcal{N}_i$ containing the $N = 30$ nearest nodes to $\mathbf{x}_i$, the node itself included. Both methods reproduce polynomials of degree up to $p$ exactly, giving a truncation error of $\mathcal{O}(s^{p+1-m})$ for an operator of order $m$.

\subsection{LABFM}\label{app:labfm}

In LABFM \citep{King2020}, the weights of~\eqref{eq:discrete_op} are expanded in a set of anisotropic basis functions (ABFs),
\begin{equation}
w_{j,i}^{D} = \mathbf{W}_{ji}^{\top}\boldsymbol{\Psi}_{i}^{D} = W_{ji}^{1}\Psi_{i,1}^{D} + W_{ji}^{2}\Psi_{i,2}^{D} + W_{ji}^{3}\Psi_{i,3}^{D} + \dots,
\end{equation}
where $\mathbf{W}_{ji}$ collects the ABFs evaluated at $\mathbf{x}_{ji}$ and $\boldsymbol{\Psi}_{i}^{D}$ is a vector of coefficients. Substituting this expansion into the moment system gives a square linear system for the coefficients,
\begin{equation}
\label{eq:labfm_system}
\mathbf{A}_{i}\boldsymbol{\Psi}_{i}^{D} = \mathbf{d}^{D}, \qquad \mathbf{A}_{i} = \sum_{j\in\mathcal{N}_i}\mathbf{X}_{ji}\otimes\mathbf{W}_{ji},
\end{equation}
where $\mathbf{X}_{ji}$ is the vector of Taylor monomials forming the columns of $\mathbf{V}_i$,
\begin{equation}
\mathbf{X}_{ji} = \left[x_{ji},\, y_{ji},\, \frac{x_{ji}^{2}}{2},\, x_{ji}y_{ji},\, \frac{y_{ji}^{2}}{2},\, \frac{x_{ji}^{3}}{6},\, \dots\right]^{\top}.
\end{equation}
For the operators used in this work, the target vector takes the form
\begin{equation}
\label{eq:labfm_targets}
\mathbf{d}^{D} =
\begin{cases}
[1, 0, 0, 0, 0, 0, \dots]^{\top} & \text{if } D = \partial_x, \\
[0, 1, 0, 0, 0, 0, \dots]^{\top} & \text{if } D = \partial_y, \\
[0, 0, 1, 0, 1, 0, \dots]^{\top} & \text{if } D = \nabla^{2}.
\end{cases}
\end{equation}
Since the constant monomial is excluded, $\mathbf{A}_{i} \in \mathbb{R}^{N_p \times N_p}$, matching the number of rows of $\mathbf{V}_i$. It is non-symmetric and is solved by LU factorisation with partial pivoting.

Each ABF is paired with one monomial of $\mathbf{X}_{ji}$. If an entry of $\mathbf{X}_{ji}$ corresponds to the multi-index $\boldsymbol{\alpha} = (\alpha_1, \alpha_2)$, the paired ABF combines a radial kernel with a product of Hermite polynomials,
\begin{equation}
\label{eq:abf}
W_{ji}^{\boldsymbol{\alpha}} = \frac{\kappa\left(\|\mathbf{x}_{ji}\|/h\right)}{\sqrt{2^{|\boldsymbol{\alpha}|}}}\, H_{\alpha_1}\!\left(\frac{x_{ji}}{h\sqrt{2}}\right) H_{\alpha_2}\!\left(\frac{y_{ji}}{h\sqrt{2}}\right),
\end{equation}
where $H_{n}$ is the physicists' Hermite polynomial of degree $n$ and $\kappa$ is the Wendland $C2$ kernel, following \citet{King2022}. The smoothing length is $h = 2.5s$. 

\subsection{RBF-FD}\label{app:rbffd}

The RBF-FD baseline uses the polyharmonic spline $\chi(r) = r^{5}$ augmented with all monomials up to degree $p$, with the zeroth-order moment included \citep{flyer2016}. The kernel has no shape parameter and requires $p \ge 2$. On the same $N = 30$ node stencils, the weights solve the system
\begin{equation}
\label{eq:rbffd_system}
\begin{bmatrix} \mathbf{K}_{i} & \mathbf{M}_{i} \\ \mathbf{M}_{i}^{\top} & \mathbf{0} \end{bmatrix}
\begin{bmatrix} \bar{\mathbf{w}}_{i}^{D} \\ \boldsymbol{\xi}_{i} \end{bmatrix}
=
\begin{bmatrix} \mathbf{c}_{i}^{D} \\ \mathbf{d}_{0}^{D} \end{bmatrix},
\end{equation}
where $(\mathbf{K}_{i})_{jj'} = \|\bar{\mathbf{x}}_{ji} - \bar{\mathbf{x}}_{j'i}\|^{5}$, the $j$-th column of $\mathbf{M}_{i}^{\top}$ is $\begin{bmatrix} 1 \\ \mathbf{X}(\bar{\mathbf{x}}_{ji}) \end{bmatrix}$, and $\mathbf{d}_{0}^{D} = \begin{bmatrix} 0 \\ \mathbf{d}^{D} \end{bmatrix}$. The entries of $\mathbf{c}_{i}^{D}$ are the continuous differentiator operator applied to the kernel at the central node, $c_{j,i}^{D} = D\,\chi(\|\mathbf{x} - \bar{\mathbf{x}}_{ji}\|)\big|_{\mathbf{x}=\mathbf{0}}$. The system is solved by dense LU and $\boldsymbol{\xi}_{i}$ is discarded. Physical weights are recovered as $w_{j,i}^{D} = \tilde{w}_{j,i}^{D}/R_i^{\,m}$, similarly to SpeND. 

\section{Experimental Details \& Additional Results}
\label{app:experiments}

All three methods run within the same Fortran solver. SpeND weights are generated by calling the trained operators from Fortran through TorchFort (with inference also performed on the CPU), which leaves the rest of the pipeline identical across methods. 
\subsection{Viscous Burgers Equation}
\label{app:burgers}

\subsubsection{Setup}
We solve the two-dimensional viscous Burgers equation in advective form,
\begin{equation}
    \partial_t \mathbf{u} + (\mathbf{u}\cdot\nabla)\mathbf{u} = \mathit{Re}^{-1} \nabla^2 \mathbf{u}, \qquad \mathbf{u} = (u, v),
\end{equation}
with $\mathit{Re} = 100$ on $\Omega = [0,1]^2$, periodic in both directions, for $t \in [0,1]$. The initial condition $u = \sin(2\pi x)$, $v = 0$ keeps $v \equiv 0$ exactly, so the solution is one-dimensional. The equation for $u$ is advanced with the full two-dimensional operators on scattered nodes. A shock forms at $x = 0.5$ and then decays. The reference solution is the Cole--Hopf series, truncated at 29 terms.

Nodes are placed at the cell centres of a Cartesian lattice with spacing $s = 1/(n+1)$, $n \in \{40, 80, 160, 320\}$. Each coordinate is then perturbed independently by $\varepsilon s\,\mathcal{U}(-\tfrac{1}{2}, \tfrac{1}{2})$ with $\varepsilon = 0.8$, with a new perturbation at each resolution. All methods use $p = 2$ operators, and SpeND obtains $\partial_y$ by applying the trained $\partial_x$ operator to stencils with the coordinates swapped. Time integration uses $4^{\text{th}}$ order Runge--Kutta. The $L_2$ error is the root-mean-square difference between the computed and reference $u$ over all nodes at the final time.

\subsubsection{Additional Results}
\begin{figure}[h]
  \begin{center}
    \begin{subfigure}{0.3\linewidth}
      \centering
      \includegraphics[width=\linewidth]{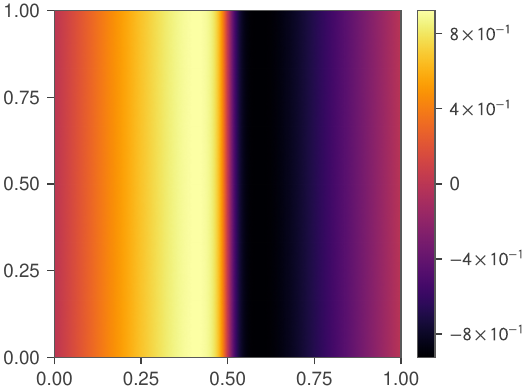}
    \end{subfigure}
    \hspace{0.02\linewidth}
    \begin{subfigure}{0.3\linewidth}
      \centering
      \includegraphics[width=\linewidth]{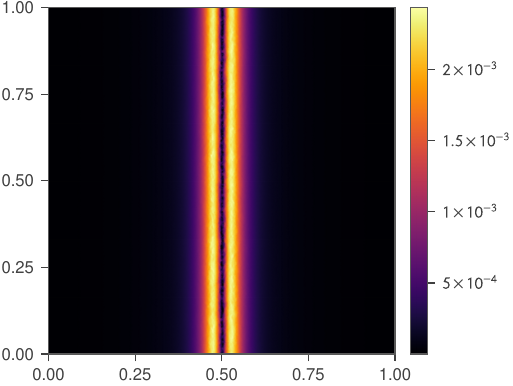}
    \end{subfigure}
    \hspace{0.02\linewidth}
    \begin{subfigure}{0.3\linewidth}
      \centering
      \includegraphics[width=\linewidth]{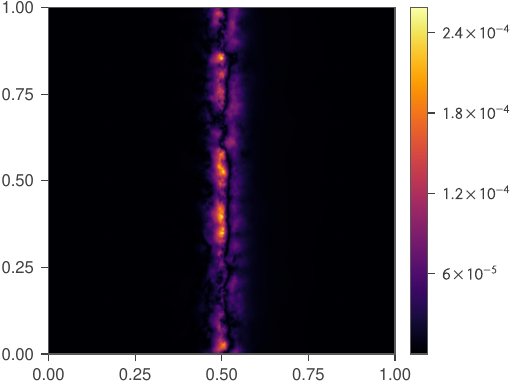}
    \end{subfigure}
    
    \vspace{0.8em}
        \begin{subfigure}{0.3\linewidth}
      \centering
      \includegraphics[width=\linewidth]{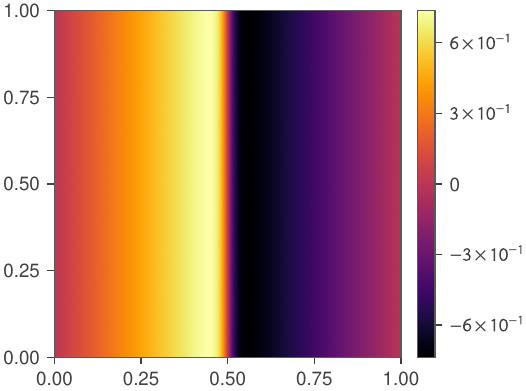}
    \end{subfigure}
    \hspace{0.02\linewidth}
    \begin{subfigure}{0.3\linewidth}
      \centering
      \includegraphics[width=\linewidth]{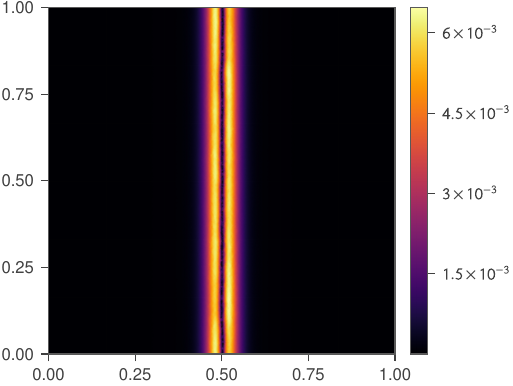}
    \end{subfigure}
    \hspace{0.02\linewidth}
    \begin{subfigure}{0.3\linewidth}
      \centering
      \includegraphics[width=\linewidth]{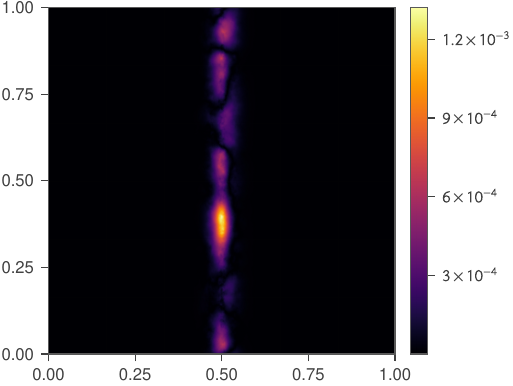}
    \end{subfigure}
    
\vspace{0.8em}
    \begin{subfigure}{0.3\linewidth}
      \centering
      \includegraphics[width=\linewidth]{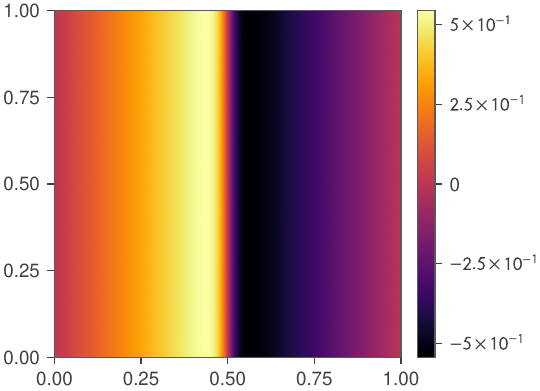}
    \end{subfigure}
    \hspace{0.02\linewidth}
    \begin{subfigure}{0.3\linewidth}
      \centering
      \includegraphics[width=\linewidth]{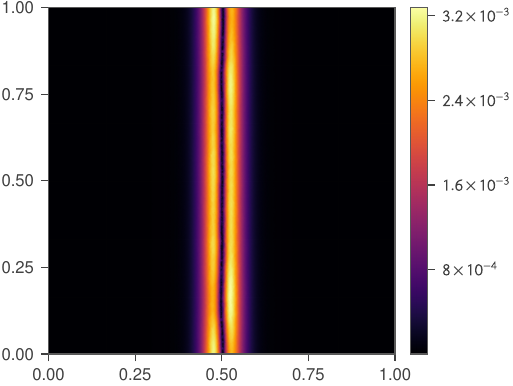}
    \end{subfigure}
    \hspace{0.02\linewidth}
    \begin{subfigure}{0.3\linewidth}
      \centering
      \includegraphics[width=\linewidth]{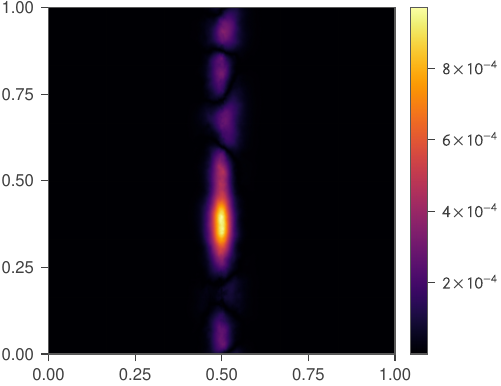}
    \end{subfigure}
    
\vspace{0.8em}
    \begin{subfigure}{0.3\linewidth}
      \centering
      \includegraphics[width=\linewidth]{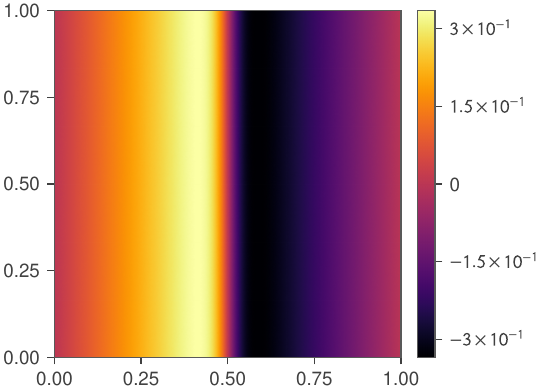}
    \end{subfigure}
    \hspace{0.02\linewidth}
    \begin{subfigure}{0.3\linewidth}
      \centering
      \includegraphics[width=\linewidth]{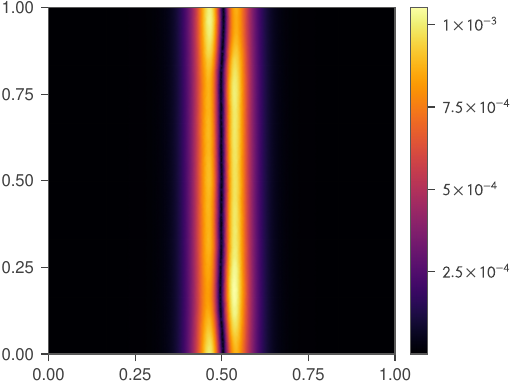}
    \end{subfigure}
    \hspace{0.02\linewidth}
    \begin{subfigure}{0.3\linewidth}
      \centering
      \includegraphics[width=\linewidth]{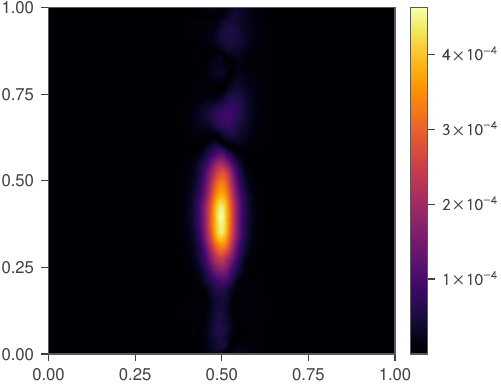}
    \end{subfigure}

  \end{center}
\caption{Viscous Burgers problem at mean node spacing $s \approx 3\times10^{-3}$. Left: analytical solution, coloured by the $u$ component of velocity. Middle and right: absolute error with respect to the analytical solution, for LABFM and SpeND, respectively. Rows correspond, from top to bottom, to $t = 0.19$, $0.39$, $0.59$, and $0.99$.}
    \label{fig:burgers_fields}
\end{figure}

Figure~\ref{fig:burgers_fields} shows the error fields of LABFM and SpeND at the finest resolution. The LABFM error is concentrated in a region around the shock and is nearly uniform in $y$. The SpeND error instead peaks on the shock itself and varies with $y$. Throughout the simulation SpeND's error remains smaller than LABFM's.

\subsection{Poisson's Equation}
\label{app:poisson}
\subsubsection{Setup}

We solve Poisson's equation
\begin{equation}
    \nabla^2 \phi = f \quad \text{in } \Omega,
\end{equation}
where $\Omega$ is the unit square with a circular obstacle of radius $r_{\mathrm{c}} = 0.2$ removed from its centre. The outer boundaries are periodic, and a Dirichlet condition is imposed on the obstacle boundary. The source term follows from the manufactured solution $\phi = \sin(6\pi x)\sin(6\pi y)$, which is periodic on the unit square, giving $f = -72\pi^2 \sin(6\pi x)\sin(6\pi y)$.

Nodes are placed at the cell centres of a Cartesian lattice with spacing $s = 1/(n+1)$, $n \in \{20, 40, 80, 160, 320, 640\}$, perturbed independently by $\varepsilon s\,\mathcal{U}(-\tfrac{1}{2}, \tfrac{1}{2})$ with $\varepsilon = 1$ in each coordinate, and those inside the obstacle are removed. The obstacle boundary carries equispaced nodes at its surface, which remain fixed while the interior is relaxed with particle shifting. All methods use $p = 3$ Laplacian operators. The sparse system is solved for all methods by BiCGSTAB with a Jacobi preconditioner. The error is the relative root-mean-square difference between the computed and exact $\phi$ over all nodes.

\subsubsection{Additional Results}
\begin{figure}[h]
  \begin{center}
    \begin{subfigure}{0.31\linewidth}
      \centering
      \includegraphics[width=\linewidth]{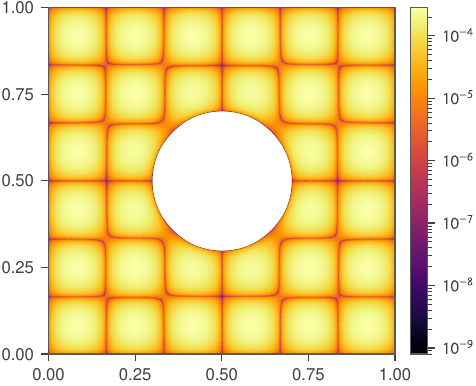}
    \end{subfigure}
    \hspace{0.015\linewidth}
    \begin{subfigure}{0.31\linewidth}
      \centering
      \includegraphics[width=\linewidth]{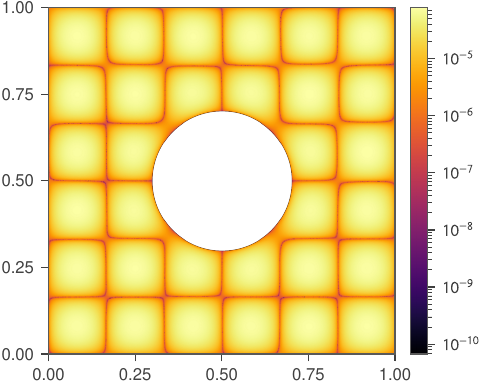}
    \end{subfigure}
    \hspace{0.015\linewidth}
    \begin{subfigure}{0.31\linewidth}
      \centering
      \includegraphics[width=\linewidth]{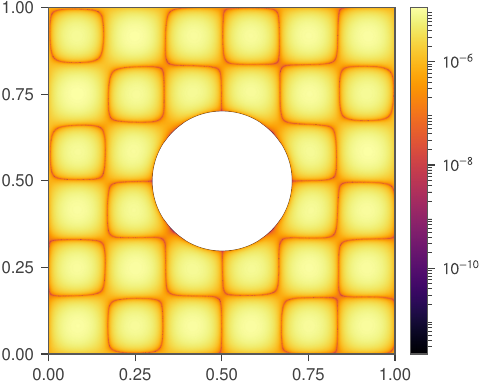}
    \end{subfigure}
    
  \end{center}
\caption{Poisson's equation with average particle spacing $s \approx 1.6\times10^{-3}$. Absolute error with respect to the analytical solution, on a log colour scale, for LABFM, RBF-FD and SpeND, from left to right.}
    \label{fig:poisson_error_map}
\end{figure}

Figure~\ref{fig:poisson_error_map} shows that the three operators produce errors with a very similar spatial structure. For all methods, the error concentrates at the crests and troughs of the manufactured solution and is smallest along the lines where $\phi$ vanishes, so each error map inherits the symmetry of $\phi$. The methods therefore differ in magnitude rather than in structure, with the maximum absolute error decreasing by roughly an order of magnitude from LABFM to RBF-FD and again from RBF-FD to SpeND.

\subsection{Weakly Compressible Navier--Stokes Equations}
\label{app:tgv}

\subsubsection{Setup}

We solve the isothermal, weakly compressible Navier--Stokes equations in conservative form,
\begin{subequations}
\label{eq:wcns}
\begin{align}
    \partial_t \rho + \nabla\cdot(\rho\mathbf{u}) &= 0, \\
    \partial_t (\rho\mathbf{u}) + \nabla\cdot(\rho\mathbf{u}\mathbf{u}) &= -Ma^{-2}\nabla\rho + \mathit{Re}^{-1}\nabla^2\mathbf{u},
\end{align}
\end{subequations}
where the system is closed by a barotropic equation of state absorbed into the momentum equation. The domain is $\Omega = [-0.5,0.5]^2$, periodic in both directions, and the flow is advanced to $t = 15$ ($t = 20$ for $s = 1/128$) with $\mathit{Re} = 100$ and $Ma = 0.1$. The reference solution is the Taylor--Green vortex,
\begin{subequations}
\label{eq:tgv_sol}
\begin{align}
    u(x,y,t) &= -\exp\!\left(-\frac{8\pi^2}{Re}\,t\right)\cos(2\pi x)\sin(2\pi y), \\
    v(x,y,t) &= \phantom{-}\exp\!\left(-\frac{8\pi^2}{Re}\,t\right)\sin(2\pi x)\cos(2\pi y),
\end{align}
\end{subequations}
evaluated at $t = 0$ for the initial velocity, with the initial density
\begin{equation}
    \rho(x,y,0) = 1 - \frac{Ma^2}{4}\left[\cos(4\pi x) + \cos(4\pi y)\right].
\end{equation}

Node sets with target spacing $s \in \{1/32, 1/64, 1/128\}$ are generated once and shared by all methods, and all methods use $p = 4$ operators. Each method is stabilised by an eighth-order hyperviscous filter applied to $\rho$, $\rho u$ and $\rho v$ after the final stage of every time step. The filter order and target damping are common to all methods. Time integration uses the four-stage, third-order low-storage Runge--Kutta scheme RK3(2)4[2R+]C of \citet{kennedy2000}, with adaptive time stepping. The error is the $L_2$ difference between computed and reference velocity magnitudes over all nodes, normalised by the reference norm at the first output time.

\subsubsection{Additional Results}
\begin{figure}[h]
  \begin{center}
    \begin{subfigure}{0.3\linewidth}
      \centering
      \includegraphics[width=\linewidth]{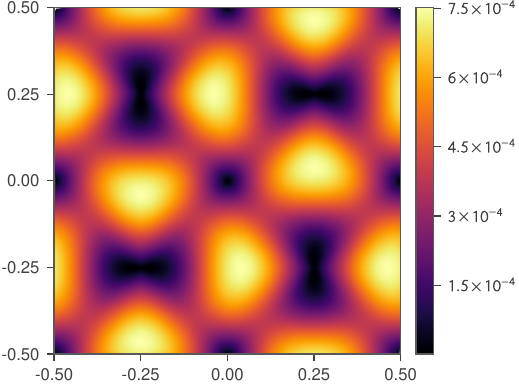}
    \end{subfigure}
    \hspace{0.03\linewidth}
    \begin{subfigure}{0.3\linewidth}
      \centering
      \includegraphics[width=\linewidth]{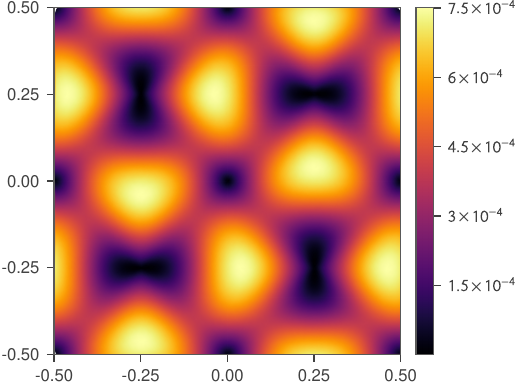}
    \end{subfigure}
    \hspace{0.03\linewidth}
    \begin{subfigure}{0.3\linewidth}
      \centering
      \includegraphics[width=\linewidth]{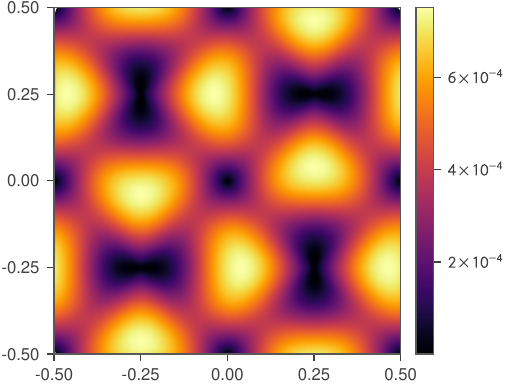}
    \end{subfigure}
    
    \vspace{0.8em}
        \begin{subfigure}{0.3\linewidth}
      \centering
      \includegraphics[width=\linewidth]{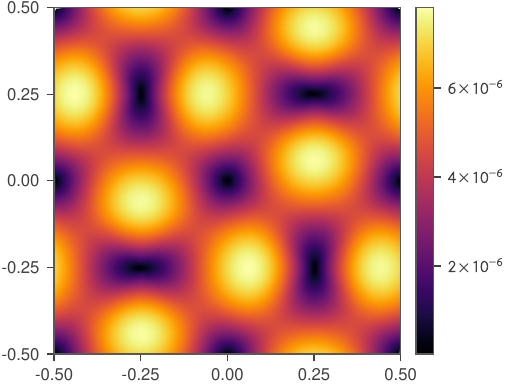}
    \end{subfigure}
    \hspace{0.03\linewidth}
    \begin{subfigure}{0.3\linewidth}
      \centering
      \includegraphics[width=\linewidth]{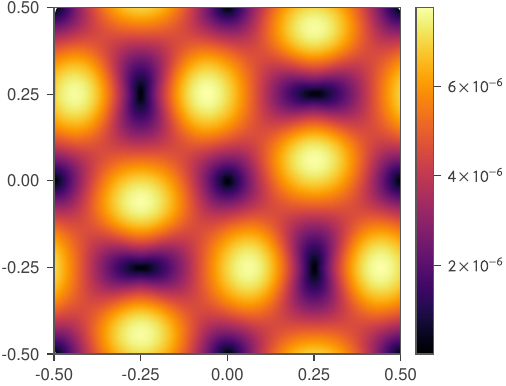}
    \end{subfigure}
    \hspace{0.03\linewidth}
    \begin{subfigure}{0.3\linewidth}
      \centering
      \includegraphics[width=\linewidth]{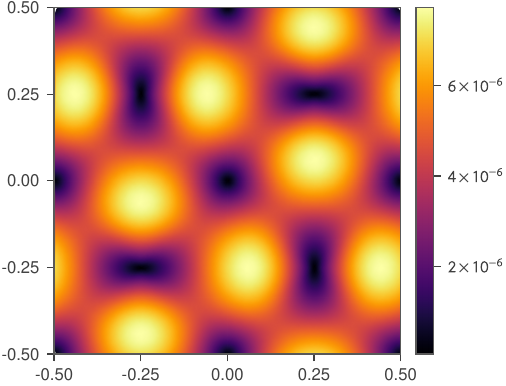}
    \end{subfigure}
    
\vspace{0.8em}

    \begin{subfigure}{0.3\linewidth}
      \centering
      \includegraphics[width=\linewidth]{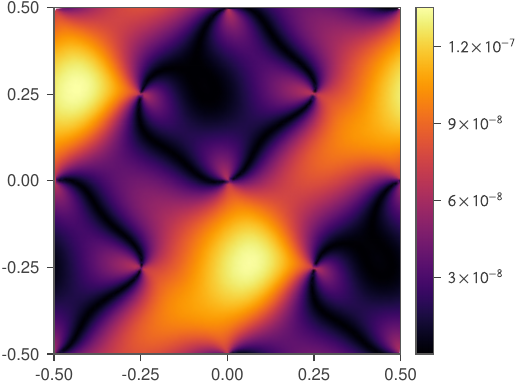}
    \end{subfigure}
    \hspace{0.03\linewidth}
    \begin{subfigure}{0.3\linewidth}
      \centering
      \includegraphics[width=\linewidth]{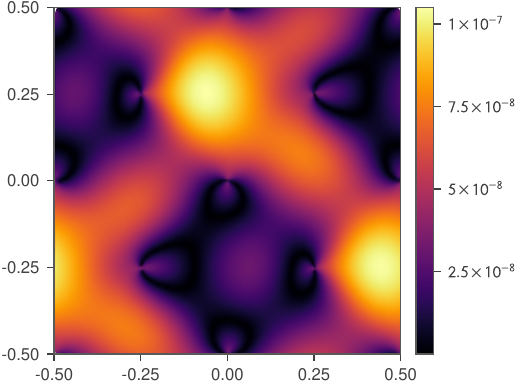}
    \end{subfigure}
    \hspace{0.03\linewidth}
    \begin{subfigure}{0.3\linewidth}
      \centering
      \includegraphics[width=\linewidth]{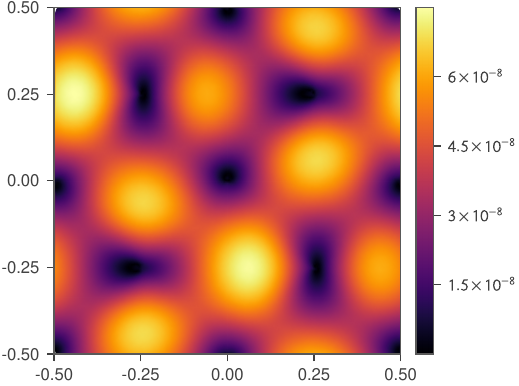}
    \end{subfigure}

  \end{center}
\caption{Taylor--Green vortex with average node spacing $s = 1/128$. Colours show the absolute error in velocity magnitude, normalised by the initial reference velocity magnitude. Columns correspond, from left to right, to LABFM, RBF-FD and SpeND, and rows, from top to bottom, to $t = 1$, $7$ and $13$.}
    \label{fig:tgv_error_time}
\end{figure}

Figure~\ref{fig:tgv_error_time} shows the spatial distribution of the error over time. Up to $t = 7$, the three methods produce visually indistinguishable error fields, and their colour bars span the same range. By $t = 13$, only SpeND preserves the symmetric structure of the Taylor--Green vortex, while the error fields of both baselines have lost the symmetry of the original problem.

\section{Limitations}
\label{sec:limitations}

All experiments are two-dimensional, and although the framework is not restricted to two dimensions, the number of moment conditions and hence the stencil size grow substantially in three dimensions, which we leave to future work. Boundary stencils are trained on truncations by a straight wall and tested only on a smooth circular obstacle, so how training and generalisation behave near complex boundaries remains to be established. The test problems are smooth and laminar at low Reynolds number; they were chosen since they admit analytical or semi-analytical solutions, isolating the error of the spatial discretisation, but as a consequence chaotic flows remain untested. Lastly, predicting the weights is more expensive than constructing them with the classical baselines, so at coarse resolutions, where this cost dominates the wall-clock time, the classical baselines reach a given error faster than SpeND; its advantage appears only once the resolution is fine enough for this one-off cost to be amortised.

\section{Ablations}
\label{app:ablation}

All ablations evaluate the $\partial_x$ and $\nabla^2$ operators on the test function of \eqref{eq:test_function}, on node sets with $n \in \{10, 20, 50, 100, 300\}$ nodes per side and spacing $s = 1/(n-1)$, obtained by disturbing a lattice by up to $s/2$ and relaxing it with $100$ particle-shifting iterations, unless specified otherwise. Within each ablation, every variant is fine-tuned for $40$ epochs from a common checkpoint and differs from the others only in the ablated setting.

\begin{figure}[h]
  \begin{center}
    \begin{subfigure}{0.3\linewidth}
      \centering
      \includegraphics[width=\linewidth]{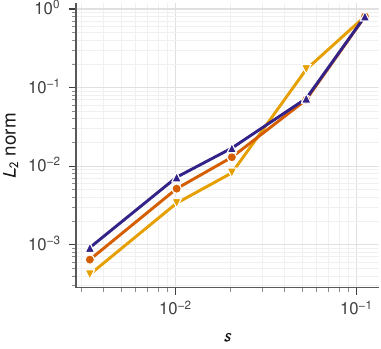}
    \end{subfigure}
    \hspace{0.015\linewidth}
    \begin{subfigure}{0.47\linewidth}
      \centering
      \includegraphics[width=\linewidth]{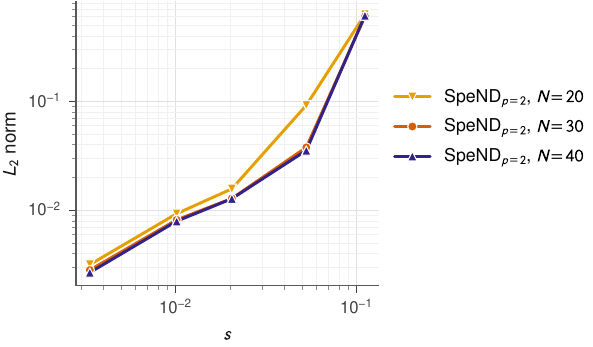}
    \end{subfigure}
  \end{center}
\caption{Effect of stencil size $N$ at $\eta = 0.7$. Relative $L_2$ error under refinement for $\partial_x$ (left) and $\nabla^2$ (right).}
  \label{fig:abl_neigh}
\end{figure}

\paragraph{Stencil size.} We train operators with $N \in \{20, 30, 40\}$ at the base configuration of Table~\ref{tab:hyperparameters}, $\eta = 0.7$, each on a corpus that differs from the others only in stencil size (Figure~\ref{fig:abl_neigh}). At the coarsest resolution, $n = 20$, reducing the stencil to $N = 20$ raises the error by roughly $2.5\times$, whereas enlarging it beyond $N = 30$ brings no further gain. Once the field is resolved, the two operators respond differently. The $\nabla^2$ error barely changes across $N$, while the $\partial_x$ error rises with $N$. A plausible explanation is the wide training band. With $\eta = 0.7$, the objective spends the additional freedom of a larger stencil on the high wavenumbers that dominate at coarse resolution rather than on the low wavenumbers that dominate once the field is resolved. The training-band ablation below supports this, since narrowing the band improves accuracy at fine resolution.

\begin{figure}[h]
  \begin{center}
    \begin{subfigure}{0.3\linewidth}
      \centering
      \includegraphics[width=\linewidth]{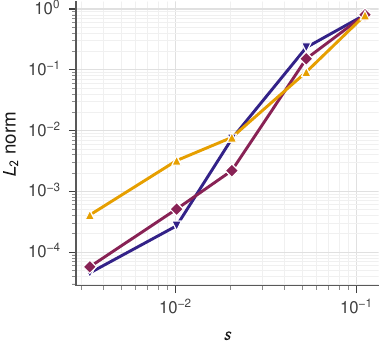}
    \end{subfigure}
    \hspace{0.015\linewidth}
    \begin{subfigure}{0.48\linewidth}
      \centering
      \includegraphics[width=\linewidth]{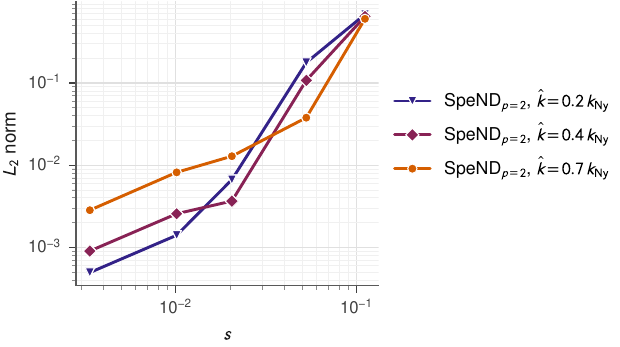}
    \end{subfigure}
  \end{center}
\caption{Effect of the band limit $\eta$ at $N = 30$. Relative $L_2$ error under refinement for $\partial_x$ (left) and $\nabla^2$ (right).}
  \label{fig:abl_band}
\end{figure}

\paragraph{Training band.} We vary the band limit $\eta \in \{0.2, 0.4, 0.7\}$ at $N = 30$ and $\epsilon = 0.1$ (Figure~\ref{fig:abl_band}). This setting has the largest effect of those ablated, and it trades accuracy between resolution regimes. At $n = 20$ the highest harmonic of \eqref{eq:test_function} lies at $0.74\,k_{\mathrm{Ny}}$, which falls the closest to the band of the $\eta = 0.7$ operators. We can see that the operators trained with $\eta=0.7$ generally perform better at coarser regions, particularly the Laplacian. At $n = 300$ every harmonic lies within all three bands and the ordering reverses, with $\eta = 0.2$ reducing the error by $8\times$ and $6\times$ relative to $\eta = 0.7$ for $\partial_x$ and $\nabla^2$, respectively. The curves cross between $n = 20$ and $n = 50$. Narrowing the band therefore improves accuracy at the wavenumbers it covers at the expense of those it excludes. The value $\eta = 0.4$ used in Section~\ref{sec:experiments} is a compromise between the two regimes.

\begin{figure}[h]
  \begin{center}
    \begin{subfigure}{0.3\linewidth}
      \centering
      \includegraphics[width=\linewidth]{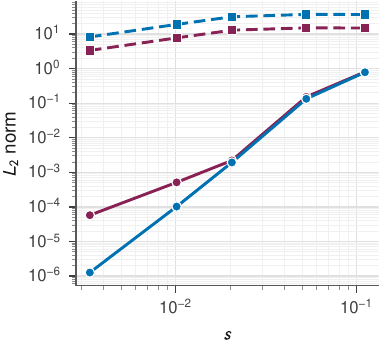}
    \end{subfigure}
    \hspace{0.015\linewidth}
    \begin{subfigure}{0.47\linewidth}
      \centering
      \includegraphics[width=\linewidth]{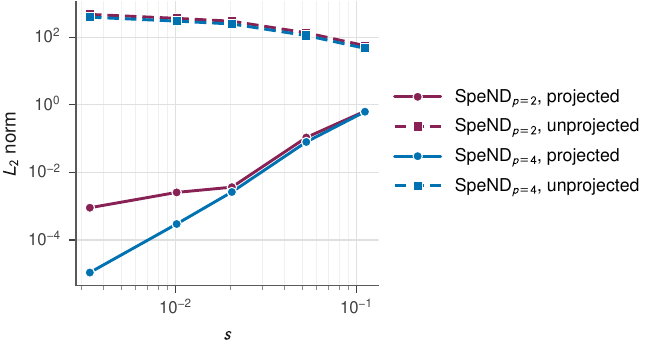}
    \end{subfigure}
  \end{center}
\caption{Effect of removing the moment projection at inference, at $p = 2$ and $p = 4$. Relative $L_2$ error under refinement for $\partial_x$ (left) and $\nabla^2$ (right).}
  \label{fig:abl_proj}
\end{figure}

\paragraph{Moment projection.} Replacing $\Pi_i$ by the identity at inference, with the network unchanged, yields relative errors between $3$ and $37$ for $\partial_x$ and between $10^2$ and $5\times10^2$ for $\nabla^2$ (Figure~\ref{fig:abl_proj}). These curves show that the moment projection is an essential aspect of the framework provided, ensuring physical consistency and, thus, formal convergence.

\begin{figure}[h]
  \begin{center}
    \begin{subfigure}{0.3\linewidth}
      \centering
      \includegraphics[width=\linewidth]{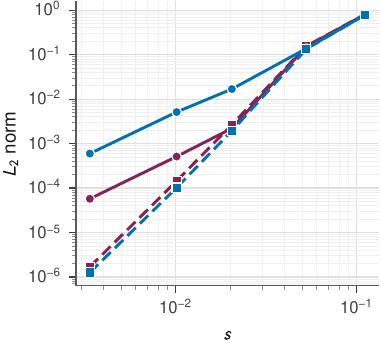}
    \end{subfigure}
    \hspace{0.015\linewidth}
    \begin{subfigure}{0.47\linewidth}
      \centering
      \includegraphics[width=\linewidth]{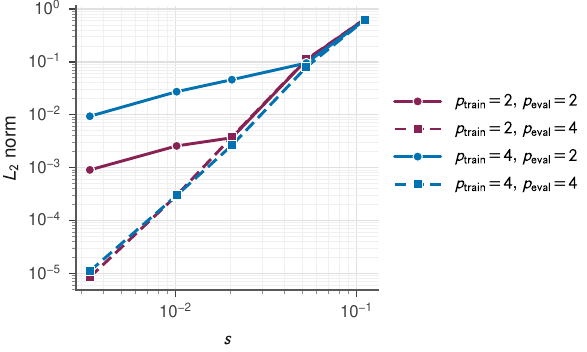}
    \end{subfigure}
  \end{center}
\caption{Networks trained at $p = 2$ and $p = 4$, each projected at $p = 2$ and $p = 4$ at inference. Relative $L_2$ error under refinement for $\partial_x$ (left) and $\nabla^2$ (right).}
  \label{fig:abl_order}
\end{figure}

\paragraph{Consistency order at inference.} Since $\bar{\mathbf{V}}_i$ is assembled at inference, a network trained under one set of moment conditions can be projected under another. We denote by $p_{\mathrm{train}}$ the consistency order imposed by the projection during training and by $p_{\mathrm{eval}}$ the order imposed at evaluation. We take the $\eta = 0.4$ network with $p_{\mathrm{train}} = 2$ and the network with $p_{\mathrm{train}} = 4$ used in Section~\ref{sec:experiments}, and evaluate each at $p_{\mathrm{eval}} \in \{2, 4\}$ (Figure~\ref{fig:abl_order}). With $p_{\mathrm{train}} = 2$ and $p_{\mathrm{eval}} = 4$, the operators recover the fourth-order convergence rates and remain close to those with $p_{\mathrm{train}} = 4$ throughout, so on this test a network can be trained at a lower order than the one imposed at evaluation. The reverse does not hold. With $p_{\mathrm{train}} = 4$ and $p_{\mathrm{eval}} = 2$, the error rises by roughly $10\times$ relative to $p_{\mathrm{train}} = p_{\mathrm{eval}} = 2$. When $p_{\mathrm{train}} = 4$, the projection fixes the third- and fourth-order moments during training, and since these set the leading error of the low-wavenumber response, the network is never trained to control it. At $p_{\mathrm{eval}} = 2$ these moments are no longer imposed, and the low-wavenumber response degrades accordingly.

\begin{figure}[h]
  \begin{center}
    \begin{subfigure}{0.3\linewidth}
      \centering
      \includegraphics[width=\linewidth]{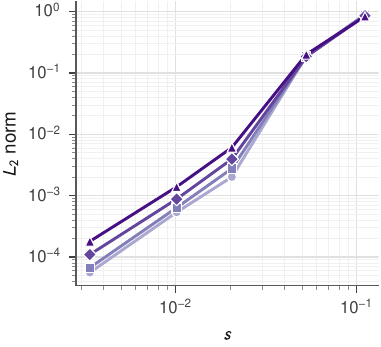}
    \end{subfigure}
    \hspace{0.015\linewidth}
    \begin{subfigure}{0.47\linewidth}
      \centering
      \includegraphics[width=\linewidth]{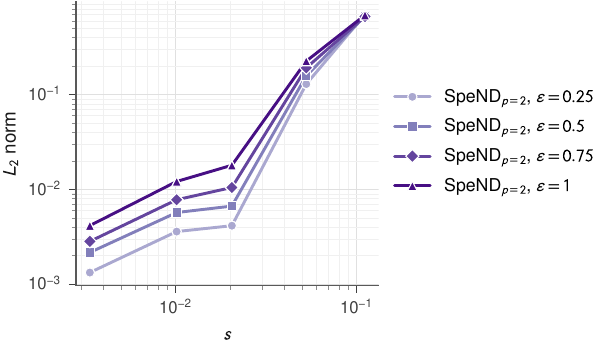}
    \end{subfigure}

    \begin{subfigure}{0.3\linewidth}
      \centering
      \includegraphics[width=\linewidth]{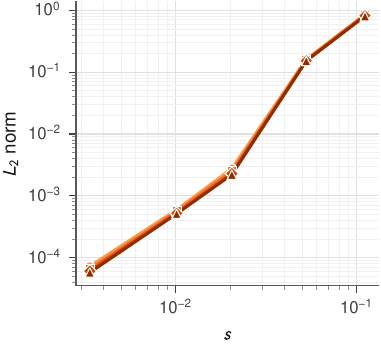}
    \end{subfigure}
    \hspace{0.015\linewidth}
    \begin{subfigure}{0.47\linewidth}
      \centering
      \includegraphics[width=\linewidth]{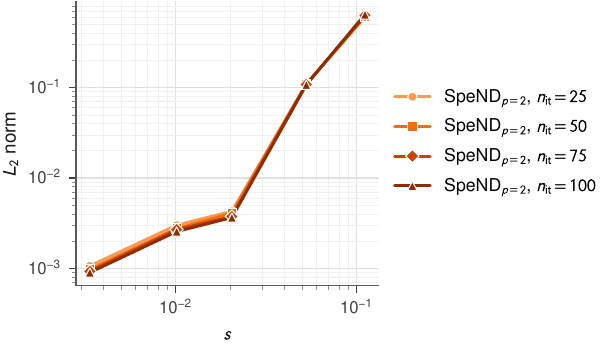}
    \end{subfigure}
  \end{center}
\caption{Effect of the node distribution on the $p = 2$ operators. Relative $L_2$ error under refinement for $\partial_x$ (left) and $\nabla^2$ (right). Top row, disturbed Cartesian lattices with disorder $\varepsilon$. Bottom row, particle-shifted node sets after the indicated number of iterations.}
  \label{fig:abl_nodes}
\end{figure}

\paragraph{Node distribution.} In place of the node sets above, we evaluate the $p = 2$ operators on two families (Figure~\ref{fig:abl_nodes}). The first consists of Cartesian lattices disturbed by $\varepsilon s\,\mathcal{U}(-\tfrac{1}{2}, \tfrac{1}{2})$ per axis with $\varepsilon \in \{0.25, 0.5, 0.75, 1\}$. The second consists of snapshots after $25$, $50$, $75$ and $100$ particle-shifting iterations from the $\varepsilon = 1$ lattice, the last being the node set used in the other ablations. SpeND retains the convergence rates of $p = 2$ on every node set, and only showing vertical shift in certain locations. On the disturbed lattices the error of both operators rises monotonically with $\varepsilon$ at every $n \geq 50$, by roughly $3\times$ from $\varepsilon = 0.25$ to $\varepsilon = 1$ at $n = 300$. Particle shifting reduces the error of both operators, with most of the gain in the first $25$ iterations. Relative to the unrelaxed $\varepsilon = 1$ lattice, these iterations lower the error at $n = 300$ by $2.5\times$ for $\partial_x$ and $3.8\times$ for $\nabla^2$, with little improvement in the further iterations.

\paragraph{Stencil-size transfer.} Since the network shares its per-point maps and pools symmetrically over the stencil, it can be applied without modification to stencils of a size other than the one it was trained on. We denote by $N_{\mathrm{train}}$ the stencil size used during training and by $N_{\mathrm{eval}}$ the stencil size at evaluation. We take the $\eta = 0.4$, $p = 2$ operators with $N_{\mathrm{train}} = 30$ and evaluate them at $N_{\mathrm{eval}} \in \{20, 30, 40\}$ on the same node sets, without retraining (Figure~\ref{fig:abl_transfer}). The output scaling $\sigma_m = \bar{s}_N^{-m}$ of Appendix~\ref{app:architecture} depends on the stencil size, and we consider two choices at evaluation. In the first, $\sigma_m$ is kept at its trained value, computed from $N_{\mathrm{train}}$. In the second, referred to as rescaled, it is recomputed from $N_{\mathrm{eval}}$. With the trained scaling, both operators converge at the tested $N_{\mathrm{eval}}$, however using $N_{\mathrm{eval}} = N_{\mathrm{train}}$ still leads to better results. Although at $N_{\mathrm{eval}}=40$ is barely distinguishable than when $N_{\mathrm{eval}}=30$ for $\partial_x$. Rescaling is considerably worse, raising the error at $n = 300$ by $17\times$ and $34\times$ for $\partial_x$ at $N_{\mathrm{eval}} = 20$ and $40$, and by $7\times$ and $8\times$ for $\nabla^2$, relative to the trained scaling. A single network can therefore serve a range of stencil sizes at a moderate loss of accuracy, provided its output scaling is left as trained.

\begin{figure}[h]
  \begin{center}
    \begin{subfigure}{0.3\linewidth}
      \centering
      \includegraphics[width=\linewidth]{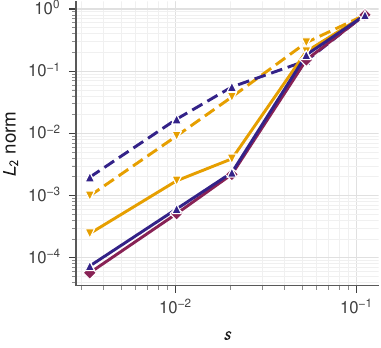}
    \end{subfigure}
    \hspace{0.015\linewidth}
    \begin{subfigure}{0.53\linewidth}
      \centering
      \includegraphics[width=\linewidth]{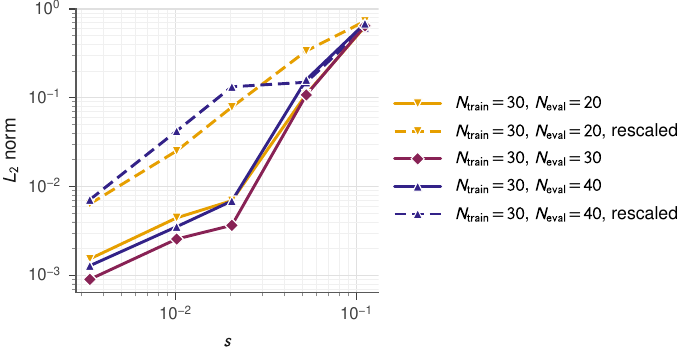}
    \end{subfigure}
  \end{center}
\caption{Operators trained at $N_{\mathrm{train}} = 30$ evaluated at $N_{\mathrm{eval}} \in \{20, 30, 40\}$ without retraining. Relative $L_2$ error under refinement for $\partial_x$ (left) and $\nabla^2$ (right). Solid lines keep the trained output scaling $\sigma_m$, and dashed lines recompute it for the evaluation stencil size.}
  \label{fig:abl_transfer}
\end{figure}

\section{Hardware \& Computational Cost}
\label{app:hardware}

Training proceeds through a chain of warm-started stages, with the first stage trained from scratch and each later stage initialised from its predecessor. The second-order operators are taken from stage~3 and the fourth-order operators from stage~4. The first two stages ran on a cluster node with a single NVIDIA L40S and eight AMD EPYC cores, and the last two on a workstation with a single NVIDIA GeForce RTX~5080 and an AMD Ryzen~7 7800X3D. Both machines use PyTorch~2.8.0 built against CUDA~12.9, with float32 arithmetic and TF32 matrix multiplication. When the trained operators are exported for use in the solvers, all parameters and arithmetic are cast to float64, so that the projection layer enforces the moment conditions to double-precision. The stages are chained to reuse earlier checkpoints and reduce compute, rather than as part of the method.

Table~\ref{tab:train_cost} reports the time spent in the training loop. The fourth-order stage is roughly $20\%$ slower per step than its second-order predecessor since the moment system grows from $N_p = 5$ to $N_p = 14$ rows, enlarging the least-squares problem solved in the projection layer.

\begin{table}[h]
\centering
\caption{Training time per stage and cumulative, in hours, for a single operator on a single GPU.}
\label{tab:train_cost}
\begin{tabular}{llc cc cc}
\toprule
 & & & \multicolumn{2}{c}{Stage} & \multicolumn{2}{c}{Cumulative} \\
\cmidrule(lr){4-5}\cmidrule(lr){6-7}
Stage & GPU & Epochs & $\partial_x$ & $\nabla^2$ & $\partial_x$ & $\nabla^2$ \\
\midrule
1. Pretraining (Adam)            & L40S     & 40 & 1.53 & 1.57 & 1.53 & 1.57 \\
2. Fine-tuning (Muon)            & L40S     & 20 & 1.22 & 1.22 & 2.75 & 2.79 \\
3. Modal response, $p = 2$       & RTX 5080 & 40 & 2.31 & 2.30 & 5.06 & 5.09 \\
4. Modal response, $p = 4$       & RTX 5080 & 40 & 2.79 & 2.72 & 7.85 & 7.81 \\
\bottomrule
\end{tabular}
\end{table}

The Burgers and Poisson experiments run on the AMD Ryzen~7 7800X3D described above using OpenMP with 16 threads, and the Taylor--Green vortex runs on the same processor using MPI with four ranks, one per physical core. All methods share the same solver implementation and differ only in the discretisation, with all solver arithmetic carried out in float64. Reported times cover the full pipeline, from node generation and neighbour search through the computation of the stencil weights to the solve itself.

\end{document}

%% file: math_commands.tex
\usepackage{amsmath,amsfonts,bm}

\def\eqref#1{equation~\ref{#1}}

\def\1{\bm{1}}

\DeclareMathAlphabet{\mathsfit}{\encodingdefault}{\sfdefault}{m}{sl}
\SetMathAlphabet{\mathsfit}{bold}{\encodingdefault}{\sfdefault}{bx}{n}

